\documentclass[twocolumn,prl,superscriptaddress,floats,nobibnotes,notitlepage,citeautoscript]{revtex4-1}

\usepackage[pdftex]{graphicx,hyperref}
\usepackage[utf8]{inputenc}
\usepackage{amsmath,bm}
\usepackage{bbold}
\usepackage{color,soul}
\usepackage{xstring}
\usepackage{wasysym}
\usepackage{xspace}
\usepackage{amssymb}
\usepackage{time}
\usepackage{bbold}
\usepackage{subfigure}
\usepackage{multirow}
\usepackage{xspace}
\usepackage{url}

\def\<{\langle}
\def\>{\rangle}
\DeclareMathOperator{\Tr}{Tr}

\newcommand{\ve}[1]{\boldsymbol{#1}}

\usepackage[dvipsnames]{xcolor}
\definecolor{Zcolour}{rgb}{0.992, 0.588, 0.22}
\definecolor{Rcolour}{rgb}{0.60, 0.0, 0.67}

\makeatother

\definecolor{DarkRed}{RGB}{193,40,40}

\def\Tr{\mathop{\mathrm{Tr}}}



\def\Tr{\mathop{\mathrm{Tr}}}

\makeindex

\begin{document}

\title{Quantum Monte Carlo at the Graphene
Quantum Hall Edge}

\author{Zhenjiu Wang}
\email{zhwang@pks.mpg.de}
\affiliation{Max-Planck-Institut für Physik komplexer Systeme, Dresden 01187, Germany}
\author{David J. Luitz}
\affiliation{Physikalisches Institut, University of Bonn, Nussallee 12, 53115 Bonn, Germany} 
\affiliation{Max-Planck-Institut für Physik komplexer Systeme, Dresden 01187, Germany} 
\author{Inti Sodemann Villadiego}
\email{sodemann@pks.mpg.de} 
\affiliation{Institut für Theoretische Physik, Universität Leipzig, D-04103, Leipzig, Germany}
\affiliation{Max-Planck-Institut für Physik komplexer Systeme, Dresden 01187, Germany}

\begin{abstract}
 We study a continuum model of the interface of graphene and vacuum in the quantum hall regime  via sign-problem-free quantum Monte Carlo, allowing us to investigate the interplay of topology and strong interactions in a graphene quantum Hall edge  
  for large system sizes. 
 We focus on the topological phase transition from the spin polarized state with symmetry protected gapless helical edges to the fully charge gapped canted-antiferromagnet state with spontaneous symmetry breaking, driven by the Zeeman energy. Our large system size simulations allow us to detail the behaviour of various quantities across this transition that are amenable to be probed experimentally, such as the spatially and energy-resolved local density of states  
 and the local compressibility. We find peculiar kinks in the branches of the edge dispersion, and also an 
 unexpected large charge susceptibility in the bulk of the canted-antiferromagnet associated with its Goldstone mode.    
\end{abstract}

\maketitle

\noindent

{\it Introduction.} The quantum Hall regime in graphene-based systems has emerged as a fantastic arena for  investigating  correlated and topological states of electrons. Progress on ingenious experimental techniques, such as compressibility measurements~\cite{feldman2012unconventional,feldman2013fractional,yang2021experimental,zibrov2018even},   
non-local magnon transmission~\cite{stepanov2018long,pierce2022thermodynamics,wei2018electrical,assouline2021excitonic,fu2021gapless,zhou2021strongmagneticfield},   
scanning-tunneling microscopy, and others~\cite{wei2017mach,kim2021edge,paul2022electrically},  
have allowed to paint a remarkably rich picture on the interplay of symmetry breaking and topology in these systems. In particular at charge neutrality   an interaction driven integer quantum Hall ferromagnet is seen in  
experiments~\cite{zhang2006landau,jiang2007quantum,Young_2012_NP,Maher_2013_NP}  
 , 
which can be driven into a spin polarized state via in-plane Zeemann coupling~\cite{Young_2014_Nature} 
and with an STO substrate~\cite{Veyrat_2020_Science}.  
The landscape of possible interaction driven states at neutrality, which likely depends on the substrate,   
still remains to be fully understood. While some experiments have been consistently interpreted by viewing the state as canted 
antiferromagnet~\cite{Young_2014_Nature,MacDonald_2014_FQH,Halperin_2013_PRB,zhou2021strongmagneticfield,paul2022electrically},  
recent STM experiments have reported a prevalence of Kekule-type valence bond solid states~\cite{Li_2019_PRB,Xiaomeng_Liu_2022,Coissard_2021}.

In this letter we investigate via sign-problem-free Quantum Monte Carlo (QMC) the proposed topological phase transition at neutrality from a canted anti-ferromagnet with gapped edges into a quantum spin Hall state with topologically protected counter-propagating modes~\cite{Lee_06_Edge}.  
To this date theoretical studies of this edge phase transition   
have been restricted to  mean field  and analytical field
theoretical studies~\cite{PRL.102.206408,Kharitonov_12_Edge,PRB.90.241410,PRB.92.165110,PRB.93.045105}, 
but  there has not been an unbiased numerical study of this transition. 

We have found several noteworthy features. First, the edge displays a clear insulator to helical metal phase transition 
by increasing the Zeeman field, as expected, but in contrast to Hartree-Fock studies\cite{PRB.92.165110},   
the charge gap at the edge opens up concomitantly with the spontaneous symmetry breaking transition in the bulk  
from spin polarized to canted-antiferromagnetic state.  
This reflects the Mermin-Wagner-type destruction of long-range order of the XY spin edge texture of 
Ref.~\onlinecite{PRB.92.165110} from quantum fluctuations as discussed in field theoretical models~\cite{PRL.97.116805,PRL.102.206408}.
Nevertheless, we observe clear kinks in the quasiparticle dispersion of the helical edge of the topological 
spin polarized state as a function of the distance to the edge, that are visible even at the metal-insulator critical 
point, which closely resemble the kinks reported in the Hartree-Forck study of Knothe and Jolicoeur~\cite{PRB.92.165110}.
In an effort to  guide future STM studies, we show how these kinks would 
appear in the spatially resolved local density of states. 
We have also found a substantially large bulk local charge susceptibility in the canted-antiferromagnetic state  
in contrast with the spin polarized state. This enhancement of the charge susceptibility can be suppressed by   
adding explicit symmetry breaking fields that  gap the goldstone modes of antiferromagnet.  This  
prediction could help guide the distinction of these correlated states in future measurements of local compressibility.

{\it Model.} We are interested in   
the half-filled   
zeroth landau level (ZLL) of graphene. 
We therefore project the Hilbert space onto the $4$ component spin/valley ZLL, 
such that there are $ 4 N_\phi $ single particle states in a torus pierced by  
$N_\phi$ flux quanta.   The fermion annihilation operators in real space are 
projected as:    
$\hat{\psi}_a(\bm{x}) = \sum_{n_k=1}^{N_{\phi}}  
\phi_{n_k} (\bm{x})   \hat{c}_{a, n_k }$ ($a=1,2,3,4$).  
Here $\hat{c}_{a, n_k }$ is the  
 canonical fermion  operator that annihilates a fermion at momentum $ k = 2\pi n_k /L_y$ and 
 flavor $a$,   
 and $ \phi_{n_k} (\bm{x}) $ is the $n_k$th single particle ZLL wave function in 
 Landau gauge ($\ve{A}(\ve{x})   = B(0,x)$). 
The model Hamiltonian reads: 
\begin{equation}\label{eq:def_H}
\hat{H} = \hat{H}_{ \text{Bulk} }  + \hat{H}_{\text{Edge} },  
\end{equation} 
where the bulk Hamiltonian is the model introduced by Kharitonov  \cite{Kharitonov_PRB_12},     
which has been successfully exploited to investigate the quantum hall regime of graphene\footnote{
see however Ref.~\cite{PRL.128.106803} for an interesting recent study beyond this model}.    
This model includes an $SU(4)$ invariant long range   
Coulomb interaction, a short range anisotropic interaction, 
as well as the Zeeman coupling:  
\begin{equation}
  \hat{H}_{\text{Bulk}} = \hat{H}_{ \text{Ani} } 
  +  \hat{H}_{\text{Coul}} +  \hat{H}_{\text{Zeeman}},       
\end{equation} 
with  

\begin{equation} \hat{H}_{ \text{Coul} }
 = \frac{ 1 }{2 }  \int_V \int_V  d \bm{x}  d^2 \bm{x}'   
\delta \hat{ \rho}(\bm{x})  V(\bm{x} - \bm{x}') \delta \hat{ \rho}(\bm{x'}) ,
\end{equation}

\noindent
  Here $ V( \bm{x} - \bm{x}'  )  \equiv   
  e^2 /\epsilon |\bm{x}-\bm{x}'|-e^2 /\epsilon \sqrt{|\bm{x}-\bm{x}'|^2+d^2}$, with $ d/2 $ the distance to a screening gate, and $\delta\hat{ \rho}(\bm{x})=\hat{ \ve{\psi}}^{\dagger} (\bm{x})  \hat{\ve{\psi}}(\bm{x}) - n_0$ , the density deviation away from half-filling, $n_0 =1/ (\pi l_B^2) $, ensuring  particle-hole symmetry. The interaction annisotropy  term read as:

\begin{equation} \hat{H}_{ \text{Ani} }
 = \frac{ 1 }{2 }  \int_V  d^2 \bm{x} 
[g_z\hat{ \tau}^2_z(\bm{x})  +g_\perp (\hat{ \tau}^2_x(\bm{x})+\hat{ \tau}^2_y(\bm{x})) ] ,
\end{equation}

\noindent where $\hat{ \tau}_i(\bm{x})=\hat{\ve{ \psi}}^{\dagger} (\bm{x})  \tau_i 
 \hat{\ve{ \psi}}(\bm{x})$, where $\tau_i$ are the Pauli matrices in valley space. The term $\hat{H}_{\text{Zeeman}} = h \int_V  d^2 \bm{x}  \hat{\ve{ \psi}}^{\dagger} (\bm{x})  \sigma^z  
 \hat{\ve{ \psi}}(\bm{x}) $  
 is the standard Zeeman coupling (with g-factor $g=2$) controlled by magnitude of the total magnetic field including its perpendicular and in-plane components. Increasing  Zeeman coupling  
 $h$  favors the spin polarized state, which is achieved once it exceeds the critical value
$ 4 \pi l_B^2 h_c / g_\perp  = 2/ \pi $.

\begin{figure}
\centering
\includegraphics[width=0.48\textwidth]{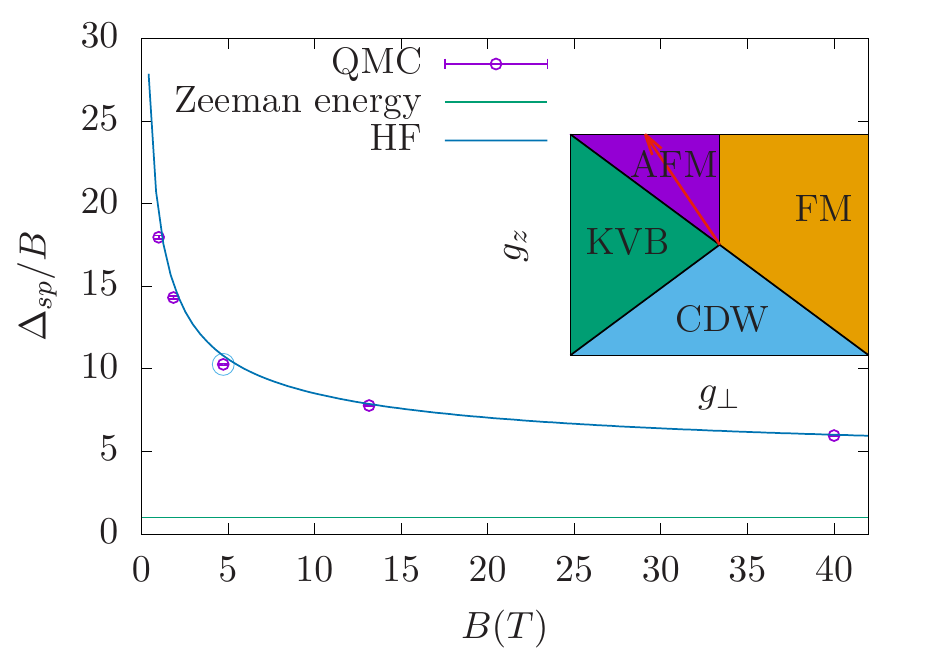}   
\caption{  
$B_\perp$ dependence of bulk single particle gap  
based on QMC simulation and HF theory.  
   Blue cycle corresponds to the  point ($B_\perp=4.7358T$) 
that we use for studying edge states.  Blue line is the gap from   
HF approximation based on Eq.~\ref{Eq:Gap_HF}.   
Inset: phase diagram of graphene quantum Hall states at 
neutrality.  
}\label{fig:Gap_B}  
\end{figure}

The phase diagram of the bulk state 
is well understood~\cite{Kharitonov_PRB_12}:   
a Kekule-valence-bond phase,   
an anti-ferromagnetic state (AFM),    
a charge-density wave state,   
as well as a ferromagnetic (FM) state as 
depicted in the inset of Fig.~\ref{fig:Gap_B}.    
Throughout this study we will  
focus on the case of   
$g_\perp < 0$, choosing $g_z = - 2 g_\perp$ 
that allows to probe the FM to AFM transition.    
Even though the charge gap remains finite across this transition,      
there is a change of the topology of the bulk state accompanied 
by an edge transition, characterised by the change of the spin 
chern number, $ C \equiv C_{ \uparrow } - C_{\downarrow} $\cite{Haldane_06},    
which is closely related to $Z_2$ time reversal invariant insulators \cite{KaneMele05,KaneMele05b}, 
albeit without time reversal symmetry. 
Here the FM state is the topologically non-trivial state with gapless helical edge modes protected   
by separate particle number conservation of spin up and down particles.   
This symmetry is spontaneously broken in the CAF state, gaping out as a result the edge states
\cite{Lee_06_Edge,Kharitonov_12_Edge}.

In order to study the edge physics,  we  
gap half  of the torus by a Kekule mass term:  
\begin{equation}\label{Eq:def_H_Kek} 
\begin{aligned}
   \hat{H}_{ \text{Edge} }   \equiv  \Delta 
 & \int_V  d^2 \bm{x}  \hat{ \psi }^{\dagger} (x, y) 
   \tau^x  \otimes \sigma^0 
  \hat{ \psi } (x, y)  s(x),  
\end{aligned}
\end{equation} 
where $s(x)$ is a smooth function that depends only on  
the 
$x$ direction which we take to be 
: 
\begin{equation}
\begin{aligned}
& s(x) \equiv \sum_{n} g(x + L_x)  \qquad n \in Z   \\ 
& g(x) \equiv \frac{1}{2} (
 \tanh [ ( x - \frac{1}{4} L_x ) / \xi ]  +
 \tanh [ ( x + \frac{1}{4} L_x ) / \xi ]   )  +  1.    
\end{aligned}  
\end{equation}      
Here $s(x)$ is nearly $0$ for $ 0<x <L_x/4  $ and  
 $ 3L_x / 4 < x <L_x $, and changing over a typical length $\xi$.  
The region where $ s(x) \approx 1 $ can be interpreted as a trivial  
vacuum, with zero spin and charge Chern numbers, and therefore serves  
as toy model to generically   capture the physical interface of  
Graphene with vacuum.  Moreover, as argued in Ref.~\onlinecite{Kharitonov_12_Edge} and Ref.~\cite{PRB.92.165110},  
the edge separating   
such vacuum regions (with $s(x)\approx 1$) from the system of interest  
(with $s(x)\approx0$ ) can  be viewed as a  
continuum approximation of the physical armchair boundaries of 
Graphene.

The applicability of  QMC simulation without sign problem 
for these terms has been previously discussed in Refs.~\cite{Ippoliti18,zwang_2020}, and we 
summarize it in Supplemental material A.  
We use the finite temperature auxiliary field 
method \cite{Blankenbecler81,White89,Assaad08_rev}  of the  algorithms for lattice fermions (ALF)-library \cite{alfcollaboration2021alf}.

\textit{ Numerical results.}  
We  take the following values for 
the  parameters that have been estimated from   
experiments~\cite{Watanabe_Young_Nature, Yacoby_2012_Science, Yacoby_2013_PRL, Halperin_2013_PRB, MacDonald_2014_FQH,  Watanabe_Young_FQH, zhou2021strongmagneticfield, hegde2022theory}       
$ h=\mu_B \sqrt{ (B_\perp^2 + B_\parallel^2) } $,   
$ g_\perp / (2\pi l_B^2) =  g_z / (4\pi l_B^2) = 10h$,     
and     
$  {e^2} / (4 \pi \epsilon l_B)  \approx  
217.63 \sqrt{B_\perp}$ where we take the  
dielectric constant $ \epsilon=4.5 $, 
 relevant for Graphene suspended in vacuum and dressed by RPA
 corrections~\cite{Sodemann_2012_RPA}.

As a first non-trivial benchmark, we will compare the bulk 
charge gap from our QMC calculations and that of Hartree-Fock theory 
(for details of HF calculation see supplementary section B):   
\begin{equation}\label{Eq:Gap_HF} 
\begin{aligned}
  \Delta_{sp}  & =    \{
 [\frac{1}{4\pi}( \widetilde{\mu_c } 
  + \mu_z - 2| \mu_\perp|) \cos\theta +h ]^2\\ 
&+[\frac{1}{4\pi}(\widetilde{\mu_c } 
  + \mu_z + 2| \mu_\perp|) \sin\theta  ]^2 \}^{1/2},  
\end{aligned}
\end{equation} 
where $ \mu_z \equiv  {g_z} / ( 2\pi l_B^2 )  $, 
 $ \mu_\perp \equiv \ {g_\perp} / ( 2\pi l_B^2 ) $ and 
 $ \widetilde{\mu}_c \equiv  
   \frac{1}{N_\phi}  \sum_{\bm{q}} 
    { f(\bm{q}) f(-\bm{q})    
   V_0 (\bm{q}) }    
   (e^2) / (\epsilon l_B ) = 0.8328  ( e^2 )  /  (\epsilon l_B)  $.     
  On the other hand, 
 $\Delta_{sp} $  
from QMC can be obtained from the asymptotic decay of the Green's 
function along the imaginary axis as follows:    
\begin{equation}
\begin{aligned}
  \frac{1}{V}  \int_V d^2  \bm{x}      
   \sum_{a} \langle 
   \hat{\psi}^{\dagger}_{a } (x, y, \tau) 
   \hat{\psi}_{a } (x, y,  0 )    \rangle   \propto 
   e^{ - \tau \Delta_{ \text{sp}} ( N_\phi ) }.     
\end{aligned}    
\end{equation} 
 The calculation is performed for 
$N_\phi = 16, 24, 32, 40$ and $48$,  with $ \beta g_\perp / (4 \pi l_B^2 ) = {N_\phi}/{2} $   
and $ 4 \pi l_B^2 \Delta_\tau / g_\perp = {4}/{N_\phi} $ (see Ref.~\cite{zwang_2020} 
for details). 
We observed very robust system size dependence of 
$\Delta_{sp}( N_\phi )$, allowing reliable extrapolation to the thermodynamic limit.
The $B_\perp$ dependence of gap in the absence of $B_\parallel$  
are plotted in Fig.~\ref{fig:Gap_B} as purple dots,   
where we see an excellent agreement with HF estimates.   
We have also performed benchmark calculations of the order parameters   
that allowed us to verify that the bulk phase transition of the CAF to FM indeed occurs at the 
critical value of $ 4 \pi l_B^2 h_c / g_\perp = 2 / \pi $.

\begin{figure*}
\centering
\includegraphics[width=0.97\textwidth]{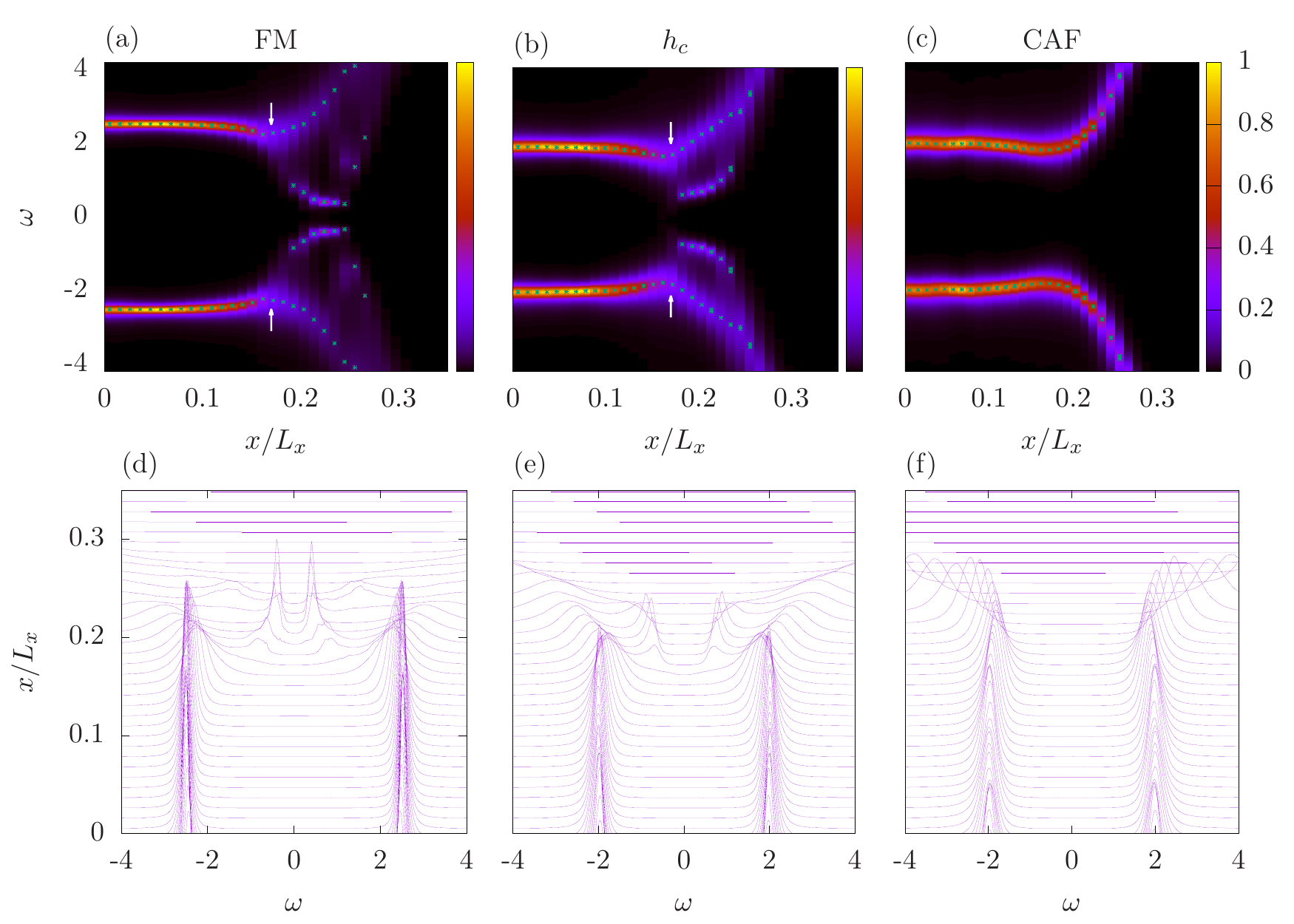} 
\caption{  
 Real space density of states  
 $ A_{\psi} ( x, \omega) $       
 for       
 $h=1.2$ ((a),(d)), $2/\pi$ ((b),(e)) and $0.2$ ((c),(f)).   
 Green dots are local maximums of  
 $ A_{\psi} ( x, \omega) $ as a function of $\omega$.   White arrows are 
 guides to the eye for the kinks of spectrum.   
}\label{fig:Green_edge} 
\end{figure*}

After having benchmarked the bulk behavior,  
we move to the main focus 
of our work, which is the study of the edge transitions.  
We pick up $B_\perp=4.7358T$ and will work in the energy unite 
where $ g_\perp / (4 \pi l_B^2) = 1 $.   
We will illustrate the behavior for three characteristic  
 values of Zeeman couplings $ h_c  =1.2$(FM), 
$2/\pi $(critical point), and $0.2$(CAF).  
The Kekule potential energy in the `vacuum' and the typical length are       
chosen as $ {\Delta}  =6$ and $\xi= 0.5$.    
One of the key    
physical observables that describes the edge  is the single particle local 
density of states: 
\begin{equation} 
\begin{aligned}
   A_{ \psi }  ( x, \omega)   = &   
  \frac{1}{Z}   \sum_{ i, j, a }  
  | \langle  i |   
   \hat{\psi}_{a }  (x ,y)    
  | j \rangle |^2  
   ( e^{ -\beta E_i } + e^{-\beta E_j } )  \\
   \times & \delta( E_i - E_j - \omega ),     
\end{aligned} 
\end{equation}
which is $y$ independent.    
$E_i$  is the $i$th  eigenvalue of Hamiltonian.           
This quantity is extracted from the imaginary time correlation    
function:  
\begin{equation}\label{Eq:def_Green}  
\begin{aligned}
   &   \sum_{a} \langle 
   \hat{\psi}^{\dagger}_{a } (x, y, \tau) 
   \hat{\psi}_{a } (x, y,  0 )   
   \rangle    
\end{aligned} 
\end{equation} 
via the stochastic maximum entropy method    \cite{Beach04,alfcollaboration2021alf}.      
For the simulation of edge states, we focus on the system size of $N_\phi=48$. 
An inverse temperature of $ \beta = 24 $ is found to be sufficient to 
converge to the  
ground state, and the Trotter step is taken as $  \Delta_\tau = 0.2 $.

The presence of the topologically protected helical edge states  
can be detected by measuring the LDOS near the edge.         
At $h=1.2$,  $A_{ \psi } (x, \omega)$ is characterised by   
 linearly dispersing edge states  
around the boundary between FM bulk and 
Kekule vacuum at $  |x| /{L_x} \approx 0.25 (0.75) $, 
as shown in Fig.~\ref{fig:Green_edge}.(a).       
Remarkably, as we see in Fig.~\ref{fig:Green_edge}, clear kinks in the spatial dependence   
of the quasi-particle dispersion appear as one moves from the spin polarized bulk towards 
the trivial edge. Such features are absent in the simplified mean field   treatment~\cite{Kharitonov_12_Edge},
but similar features were found in the more systematic Hartree-Fock  
study of Ref.~\cite{PRB.92.165110}  
Similar behavior exists at the FM-CAF critical point $h_c=2/\pi$,  
even though the quasiparticle peaks are much broader in this case as seen in panels (d,e)  
of Fig.~\ref{fig:Green_edge}. On the other hand, we have found that at the critical point 
the electron quasiparticle gap at the edge vanishes in the thermodynamic limit,  as further 
discussed in the Supplementary section C.  This contrasts with the Hartree-Fock analysis  
of  Ref.~\cite{PRB.92.165110}, but is consistent with Mermin-Wagner absence of long-range order 
for the XY spin projection for this helical Luttinger liquid as discussed   in~\cite{PRL.97.116805,PRL.102.206408}.   
Finally, at $h=0.2$,  $A_{ \psi } (x, \omega)$ shows clearly gaped   
behavior at the edge due to the spontaneously broken   
$\sigma_z$ conservation in the bulk,  and also no strong kink-like features are seen 
in the edge quasiparticle spectrum.  
Another interesting finding is that  
the broadening of quasi-particle peaks  
originates primarily from the   
long range Coulomb interaction, and not from the short distance valley dependent interactions. 
We illustrate this in detail in the Supplementary Section D, where we show the sharp 
quasiparticle peaks that would appear in a model without the long range part of the 
Coulomb interactions.

\begin{figure}
\centering
\includegraphics[width=0.48\textwidth]{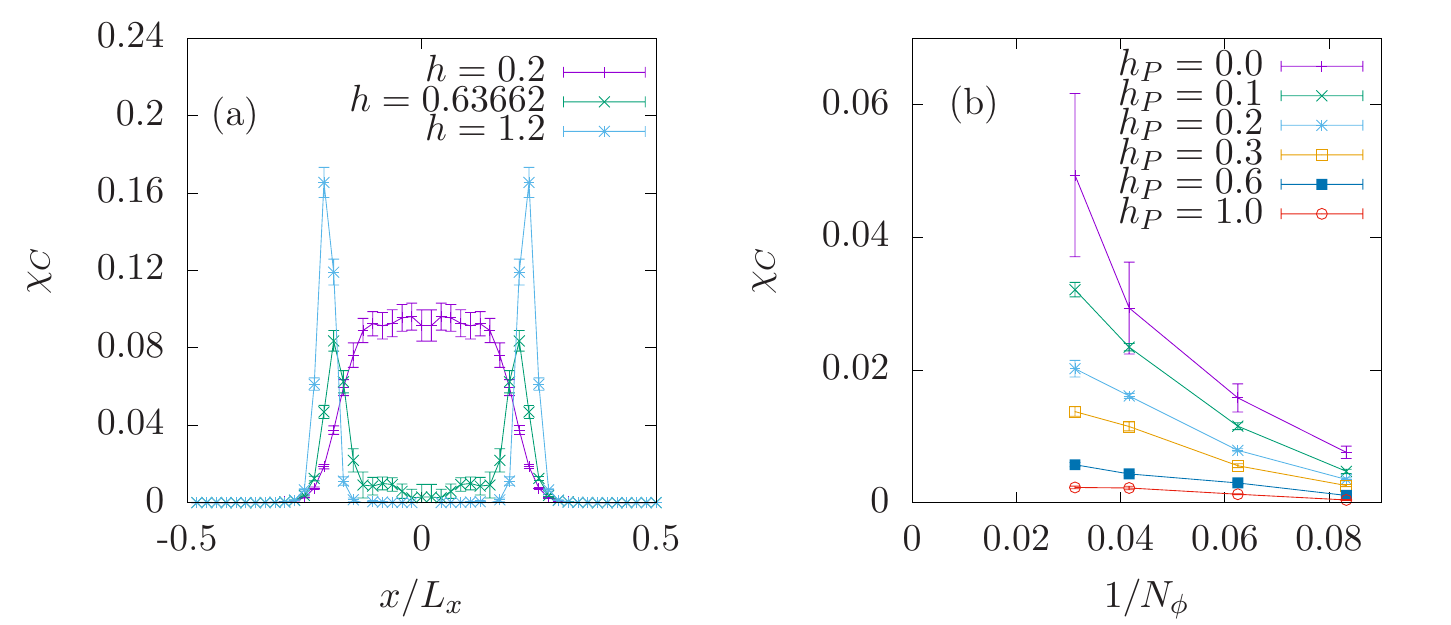}  
\caption{  
 (a). Real space dependence of $\chi_C (x)$ at $h=0.2, 0.63662$ and $1.2$ 
  for $N_\phi=48$. 
      We took  the vacuum potential $\Delta=6$. 
 (b). $N_\phi$ dependence of spatially averaged $ \chi_C $ at $h=0.2$, for $\Delta=0$.   
 We took $  \beta   = N_\phi $ and 
 the trotter step  as  $  \Delta_\tau= 8/N_\phi $.    
}\label{fig:Charge_sus} 
\end{figure}

We also determined the local charge susceptibility: 
\begin{equation}\label{Eq:charge_sus}
\begin{aligned}
   & \chi_{  \text{charge} } ( x )  \\ 
   =  & 
   \int_0^{ \beta }  d t  \langle  
   \hat{\psi}^{\dagger} (x, y, \tau)   \hat{\psi}  (x, y, \tau)  
   \hat{\psi}^{\dagger} (x, y, 0 )     \hat{\psi}  (x, y, 0 )    
   \rangle,     
\end{aligned}    
\end{equation}

$ \chi_{\text{charge}} $  displays clear peaks at the edge  for the  
FM phase and all the way to the critical point as shown  
in Fig.~\ref{fig:Charge_sus}(a),      
reflecting the compressible nature of the gapless helical edge, as expected. 
More remarkably, we have found a large local charge compressibility of the bulk of the 
CAF state, as shown in Fig.~\ref{fig:Charge_sus}. 
We have verified that this large susceptibility is also  
present in the uniform case in the absence of the edge potential.   
As a cross check,  we consider a periodic system ( $\Delta=0$ ) in  presence of a finite 
pinning potential $ H_{\text{Pin}} = h_P  \int d^2 \bm{x}  \hat{ \psi }^{\dagger} (x, y)  
 \tau^z  \otimes \sigma^z    \hat{ \psi } (x, y) $. 
We have found that  $ \chi_{\text{charge}} $ is suppressed, as shown in
Fig.~\ref{fig:Charge_sus}(b),   
demonstrating that this enhanced local charge susceptibility arises from the spontaneous 
symmetry breaking of CAF and its associated Goldstone mode.

\textit{Discussion. } We have investigated via a large scale sign-problem-free QMC technique a topological phase transition realized in the quantum Hall regime of graphene at neutrality. This is a transition from a spin polarized (FM) state with counter-propagating helical edge modes, protected by the spin conservation and a non-trivial bulk spin Chern number, into a canted-antiferromagnetic that spontaneously breaks spin conservation, accompanied by a concomitant gapping of the helical edge modes. We have computed the electron local density of states, which is a quantity amenable to be probed in scanning tunneling microscopy.    
We have seen that the quasiparticle dispersions of the FM state display non-monotonicity and 
and kink-like features that are not present in the simplified mean field treatment of 
Ref.~\cite{Kharitonov_12_Edge}, but are reminiscent of those found in more systematic 
Hartree-Fock models of the edge~\cite{PRB.92.165110}.  
We have also studied the local charge susceptibility and demonstrated that the edge 
of the spin polarized state  
remains substantially compressible all the way to the critical point, beyond which a full charge gap appears in both the edge and the bulk. 
Interestingly, we have found that the CAF state has a large local bulk charge susceptibility 
that can be suppressed by pinning the Goldstone mode.   
These charge susceptibilities are amenable to be probed by local compressibility measurements.

{\begin{acknowledgments} 
ZW would like to thank Fakher F. Assaad and Peng Rao for helpful discussions.    
The authors gratefully acknowledge the Gauss Centre for Supercomputing e.V. (www.gauss-centre.eu) for funding this project by providing computing time on the GCS Supercomputer SUPERMUC-NG at Leibniz Supercomputing Centre (www.lrz.de).  DJL acknowledges support by the DFG through SFB
1143 (project-id 247310070) and the cluster of excellence
ML4Q (EXC2004, project-id 390534769).
\end{acknowledgments}}

\bibliography{zwang}

\begin{thebibliography}{50}%
\makeatletter
\providecommand \@ifxundefined [1]{%
 \@ifx{#1\undefined}
}%
\providecommand \@ifnum [1]{%
 \ifnum #1\expandafter \@firstoftwo
 \else \expandafter \@secondoftwo
 \fi
}%
\providecommand \@ifx [1]{%
 \ifx #1\expandafter \@firstoftwo
 \else \expandafter \@secondoftwo
 \fi
}%
\providecommand \natexlab [1]{#1}%
\providecommand \enquote  [1]{``#1''}%
\providecommand \bibnamefont  [1]{#1}%
\providecommand \bibfnamefont [1]{#1}%
\providecommand \citenamefont [1]{#1}%
\providecommand \href@noop [0]{\@secondoftwo}%
\providecommand \href [0]{\begingroup \@sanitize@url \@href}%
\providecommand \@href[1]{\@@startlink{#1}\@@href}%
\providecommand \@@href[1]{\endgroup#1\@@endlink}%
\providecommand \@sanitize@url [0]{\catcode `\\12\catcode `\$12\catcode
  `\&12\catcode `\#12\catcode `\^12\catcode `\_12\catcode `\%12\relax}%
\providecommand \@@startlink[1]{}%
\providecommand \@@endlink[0]{}%
\providecommand \url  [0]{\begingroup\@sanitize@url \@url }%
\providecommand \@url [1]{\endgroup\@href {#1}{\urlprefix }}%
\providecommand \urlprefix  [0]{URL }%
\providecommand \Eprint [0]{\href }%
\providecommand \doibase [0]{http://dx.doi.org/}%
\providecommand \selectlanguage [0]{\@gobble}%
\providecommand \bibinfo  [0]{\@secondoftwo}%
\providecommand \bibfield  [0]{\@secondoftwo}%
\providecommand \translation [1]{[#1]}%
\providecommand \BibitemOpen [0]{}%
\providecommand \bibitemStop [0]{}%
\providecommand \bibitemNoStop [0]{.\EOS\space}%
\providecommand \EOS [0]{\spacefactor3000\relax}%
\providecommand \BibitemShut  [1]{\csname bibitem#1\endcsname}%
\let\auto@bib@innerbib\@empty
\bibitem [{\citenamefont {Feldman}\ \emph
  {et~al.}(2012{\natexlab{a}})\citenamefont {Feldman}, \citenamefont {Krauss},
  \citenamefont {Smet},\ and\ \citenamefont
  {Yacoby}}]{feldman2012unconventional}%
  \BibitemOpen
  \bibfield  {author} {\bibinfo {author} {\bibfnamefont {B.~E.}\ \bibnamefont
  {Feldman}}, \bibinfo {author} {\bibfnamefont {B.}~\bibnamefont {Krauss}},
  \bibinfo {author} {\bibfnamefont {J.~H.}\ \bibnamefont {Smet}}, \ and\
  \bibinfo {author} {\bibfnamefont {A.}~\bibnamefont {Yacoby}},\ }\href
  {\doibase 10.1126/science.1224784} {\bibfield  {journal} {\bibinfo  {journal}
  {Science}\ }\textbf {\bibinfo {volume} {337}},\ \bibinfo {pages} {1196}
  (\bibinfo {year} {2012}{\natexlab{a}})},\ \Eprint
  {http://arxiv.org/abs/https://www.science.org/doi/pdf/10.1126/science.1224784}
  {https://www.science.org/doi/pdf/10.1126/science.1224784} \BibitemShut
  {NoStop}%
\bibitem [{\citenamefont {Feldman}\ \emph
  {et~al.}(2013{\natexlab{a}})\citenamefont {Feldman}, \citenamefont {Levin},
  \citenamefont {Krauss}, \citenamefont {Abanin}, \citenamefont {Halperin},
  \citenamefont {Smet},\ and\ \citenamefont {Yacoby}}]{feldman2013fractional}%
  \BibitemOpen
  \bibfield  {author} {\bibinfo {author} {\bibfnamefont {B.~E.}\ \bibnamefont
  {Feldman}}, \bibinfo {author} {\bibfnamefont {A.~J.}\ \bibnamefont {Levin}},
  \bibinfo {author} {\bibfnamefont {B.}~\bibnamefont {Krauss}}, \bibinfo
  {author} {\bibfnamefont {D.~A.}\ \bibnamefont {Abanin}}, \bibinfo {author}
  {\bibfnamefont {B.~I.}\ \bibnamefont {Halperin}}, \bibinfo {author}
  {\bibfnamefont {J.~H.}\ \bibnamefont {Smet}}, \ and\ \bibinfo {author}
  {\bibfnamefont {A.}~\bibnamefont {Yacoby}},\ }\href {\doibase
  10.1103/PhysRevLett.111.076802} {\bibfield  {journal} {\bibinfo  {journal}
  {Phys. Rev. Lett.}\ }\textbf {\bibinfo {volume} {111}},\ \bibinfo {pages}
  {076802} (\bibinfo {year} {2013}{\natexlab{a}})}\BibitemShut {NoStop}%
\bibitem [{\citenamefont {Yang}\ \emph {et~al.}(2021)\citenamefont {Yang},
  \citenamefont {Zibrov}, \citenamefont {Bai}, \citenamefont {Taniguchi},
  \citenamefont {Watanabe}, \citenamefont {Zaletel},\ and\ \citenamefont
  {Young}}]{yang2021experimental}%
  \BibitemOpen
  \bibfield  {author} {\bibinfo {author} {\bibfnamefont {F.}~\bibnamefont
  {Yang}}, \bibinfo {author} {\bibfnamefont {A.~A.}\ \bibnamefont {Zibrov}},
  \bibinfo {author} {\bibfnamefont {R.}~\bibnamefont {Bai}}, \bibinfo {author}
  {\bibfnamefont {T.}~\bibnamefont {Taniguchi}}, \bibinfo {author}
  {\bibfnamefont {K.}~\bibnamefont {Watanabe}}, \bibinfo {author}
  {\bibfnamefont {M.~P.}\ \bibnamefont {Zaletel}}, \ and\ \bibinfo {author}
  {\bibfnamefont {A.~F.}\ \bibnamefont {Young}},\ }\href {\doibase
  10.1103/PhysRevLett.126.156802} {\bibfield  {journal} {\bibinfo  {journal}
  {Phys. Rev. Lett.}\ }\textbf {\bibinfo {volume} {126}},\ \bibinfo {pages}
  {156802} (\bibinfo {year} {2021})}\BibitemShut {NoStop}%
\bibitem [{\citenamefont {Zibrov}\ \emph
  {et~al.}(2018{\natexlab{a}})\citenamefont {Zibrov}, \citenamefont {Spanton},
  \citenamefont {Zhou}, \citenamefont {Kometter}, \citenamefont {Taniguchi},
  \citenamefont {Watanabe},\ and\ \citenamefont {Young}}]{zibrov2018even}%
  \BibitemOpen
  \bibfield  {author} {\bibinfo {author} {\bibfnamefont {A.~A.}\ \bibnamefont
  {Zibrov}}, \bibinfo {author} {\bibfnamefont {E.~M.}\ \bibnamefont {Spanton}},
  \bibinfo {author} {\bibfnamefont {H.}~\bibnamefont {Zhou}}, \bibinfo {author}
  {\bibfnamefont {C.}~\bibnamefont {Kometter}}, \bibinfo {author}
  {\bibfnamefont {T.}~\bibnamefont {Taniguchi}}, \bibinfo {author}
  {\bibfnamefont {K.}~\bibnamefont {Watanabe}}, \ and\ \bibinfo {author}
  {\bibfnamefont {A.~F.}\ \bibnamefont {Young}},\ }\href {\doibase
  10.1038/s41567-018-0190-0} {\bibfield  {journal} {\bibinfo  {journal} {Nature
  Physics}\ }\textbf {\bibinfo {volume} {14}},\ \bibinfo {pages} {930}
  (\bibinfo {year} {2018}{\natexlab{a}})}\BibitemShut {NoStop}%
\bibitem [{\citenamefont {Stepanov}\ \emph {et~al.}(2018)\citenamefont
  {Stepanov}, \citenamefont {Che}, \citenamefont {Shcherbakov}, \citenamefont
  {Yang}, \citenamefont {Chen}, \citenamefont {Thilahar}, \citenamefont
  {Voigt}, \citenamefont {Bockrath}, \citenamefont {Smirnov}, \citenamefont
  {Watanabe}, \citenamefont {Taniguchi}, \citenamefont {Lake}, \citenamefont
  {Barlas}, \citenamefont {MacDonald},\ and\ \citenamefont
  {Lau}}]{stepanov2018long}%
  \BibitemOpen
  \bibfield  {author} {\bibinfo {author} {\bibfnamefont {P.}~\bibnamefont
  {Stepanov}}, \bibinfo {author} {\bibfnamefont {S.}~\bibnamefont {Che}},
  \bibinfo {author} {\bibfnamefont {D.}~\bibnamefont {Shcherbakov}}, \bibinfo
  {author} {\bibfnamefont {J.}~\bibnamefont {Yang}}, \bibinfo {author}
  {\bibfnamefont {R.}~\bibnamefont {Chen}}, \bibinfo {author} {\bibfnamefont
  {K.}~\bibnamefont {Thilahar}}, \bibinfo {author} {\bibfnamefont
  {G.}~\bibnamefont {Voigt}}, \bibinfo {author} {\bibfnamefont {M.~W.}\
  \bibnamefont {Bockrath}}, \bibinfo {author} {\bibfnamefont {D.}~\bibnamefont
  {Smirnov}}, \bibinfo {author} {\bibfnamefont {K.}~\bibnamefont {Watanabe}},
  \bibinfo {author} {\bibfnamefont {T.}~\bibnamefont {Taniguchi}}, \bibinfo
  {author} {\bibfnamefont {R.~K.}\ \bibnamefont {Lake}}, \bibinfo {author}
  {\bibfnamefont {Y.}~\bibnamefont {Barlas}}, \bibinfo {author} {\bibfnamefont
  {A.~H.}\ \bibnamefont {MacDonald}}, \ and\ \bibinfo {author} {\bibfnamefont
  {C.~N.}\ \bibnamefont {Lau}},\ }\href {\doibase 10.1038/s41567-018-0161-5}
  {\bibfield  {journal} {\bibinfo  {journal} {Nature Physics}\ }\textbf
  {\bibinfo {volume} {14}},\ \bibinfo {pages} {907} (\bibinfo {year}
  {2018})}\BibitemShut {NoStop}%
\bibitem [{\citenamefont {Pierce}\ \emph {et~al.}(2022)\citenamefont {Pierce},
  \citenamefont {Xie}, \citenamefont {Lee}, \citenamefont {Forrester},
  \citenamefont {Wei}, \citenamefont {Watanabe}, \citenamefont {Taniguchi},
  \citenamefont {Halperin},\ and\ \citenamefont
  {Yacoby}}]{pierce2022thermodynamics}%
  \BibitemOpen
  \bibfield  {author} {\bibinfo {author} {\bibfnamefont {A.~T.}\ \bibnamefont
  {Pierce}}, \bibinfo {author} {\bibfnamefont {Y.}~\bibnamefont {Xie}},
  \bibinfo {author} {\bibfnamefont {S.~H.}\ \bibnamefont {Lee}}, \bibinfo
  {author} {\bibfnamefont {P.~R.}\ \bibnamefont {Forrester}}, \bibinfo {author}
  {\bibfnamefont {D.~S.}\ \bibnamefont {Wei}}, \bibinfo {author} {\bibfnamefont
  {K.}~\bibnamefont {Watanabe}}, \bibinfo {author} {\bibfnamefont
  {T.}~\bibnamefont {Taniguchi}}, \bibinfo {author} {\bibfnamefont {B.~I.}\
  \bibnamefont {Halperin}}, \ and\ \bibinfo {author} {\bibfnamefont
  {A.}~\bibnamefont {Yacoby}},\ }\href {\doibase 10.1038/s41567-021-01421-x}
  {\bibfield  {journal} {\bibinfo  {journal} {Nature Physics}\ }\textbf
  {\bibinfo {volume} {18}},\ \bibinfo {pages} {37} (\bibinfo {year}
  {2022})}\BibitemShut {NoStop}%
\bibitem [{\citenamefont {Wei}\ \emph {et~al.}(2018)\citenamefont {Wei},
  \citenamefont {van~der Sar}, \citenamefont {Lee}, \citenamefont {Watanabe},
  \citenamefont {Taniguchi}, \citenamefont {Halperin},\ and\ \citenamefont
  {Yacoby}}]{wei2018electrical}%
  \BibitemOpen
  \bibfield  {author} {\bibinfo {author} {\bibfnamefont {D.~S.}\ \bibnamefont
  {Wei}}, \bibinfo {author} {\bibfnamefont {T.}~\bibnamefont {van~der Sar}},
  \bibinfo {author} {\bibfnamefont {S.~H.}\ \bibnamefont {Lee}}, \bibinfo
  {author} {\bibfnamefont {K.}~\bibnamefont {Watanabe}}, \bibinfo {author}
  {\bibfnamefont {T.}~\bibnamefont {Taniguchi}}, \bibinfo {author}
  {\bibfnamefont {B.~I.}\ \bibnamefont {Halperin}}, \ and\ \bibinfo {author}
  {\bibfnamefont {A.}~\bibnamefont {Yacoby}},\ }\href {\doibase
  10.1126/science.aar4061} {\bibfield  {journal} {\bibinfo  {journal}
  {Science}\ }\textbf {\bibinfo {volume} {362}},\ \bibinfo {pages} {229}
  (\bibinfo {year} {2018})},\ \Eprint
  {http://arxiv.org/abs/https://www.science.org/doi/pdf/10.1126/science.aar4061}
  {https://www.science.org/doi/pdf/10.1126/science.aar4061} \BibitemShut
  {NoStop}%
\bibitem [{\citenamefont {Assouline}\ \emph {et~al.}(2021)\citenamefont
  {Assouline}, \citenamefont {Jo}, \citenamefont {Brasseur}, \citenamefont
  {Watanabe}, \citenamefont {Taniguchi}, \citenamefont {Jolicoeur},
  \citenamefont {Glattli}, \citenamefont {Kumada}, \citenamefont {Roche},
  \citenamefont {Parmentier},\ and\ \citenamefont
  {Roulleau}}]{assouline2021excitonic}%
  \BibitemOpen
  \bibfield  {author} {\bibinfo {author} {\bibfnamefont {A.}~\bibnamefont
  {Assouline}}, \bibinfo {author} {\bibfnamefont {M.}~\bibnamefont {Jo}},
  \bibinfo {author} {\bibfnamefont {P.}~\bibnamefont {Brasseur}}, \bibinfo
  {author} {\bibfnamefont {K.}~\bibnamefont {Watanabe}}, \bibinfo {author}
  {\bibfnamefont {T.}~\bibnamefont {Taniguchi}}, \bibinfo {author}
  {\bibfnamefont {T.}~\bibnamefont {Jolicoeur}}, \bibinfo {author}
  {\bibfnamefont {D.~C.}\ \bibnamefont {Glattli}}, \bibinfo {author}
  {\bibfnamefont {N.}~\bibnamefont {Kumada}}, \bibinfo {author} {\bibfnamefont
  {P.}~\bibnamefont {Roche}}, \bibinfo {author} {\bibfnamefont {F.~D.}\
  \bibnamefont {Parmentier}}, \ and\ \bibinfo {author} {\bibfnamefont
  {P.}~\bibnamefont {Roulleau}},\ }\href {\doibase 10.1038/s41567-021-01411-z}
  {\bibfield  {journal} {\bibinfo  {journal} {Nature Physics}\ }\textbf
  {\bibinfo {volume} {17}},\ \bibinfo {pages} {1369} (\bibinfo {year}
  {2021})}\BibitemShut {NoStop}%
\bibitem [{\citenamefont {Fu}\ \emph {et~al.}(2021)\citenamefont {Fu},
  \citenamefont {Huang}, \citenamefont {Watanabe}, \citenamefont {Taniguchi},\
  and\ \citenamefont {Zhu}}]{fu2021gapless}%
  \BibitemOpen
  \bibfield  {author} {\bibinfo {author} {\bibfnamefont {H.}~\bibnamefont
  {Fu}}, \bibinfo {author} {\bibfnamefont {K.}~\bibnamefont {Huang}}, \bibinfo
  {author} {\bibfnamefont {K.}~\bibnamefont {Watanabe}}, \bibinfo {author}
  {\bibfnamefont {T.}~\bibnamefont {Taniguchi}}, \ and\ \bibinfo {author}
  {\bibfnamefont {J.}~\bibnamefont {Zhu}},\ }\href {\doibase
  10.1103/PhysRevX.11.021012} {\bibfield  {journal} {\bibinfo  {journal} {Phys.
  Rev. X}\ }\textbf {\bibinfo {volume} {11}},\ \bibinfo {pages} {021012}
  (\bibinfo {year} {2021})}\BibitemShut {NoStop}%
\bibitem [{\citenamefont {Zhou}\ \emph {et~al.}(2021)\citenamefont {Zhou},
  \citenamefont {Huang}, \citenamefont {Wei}, \citenamefont {Taniguchi},
  \citenamefont {Watanabe}, \citenamefont {Zaletel}, \citenamefont {Papić},
  \citenamefont {MacDonald},\ and\ \citenamefont
  {Young}}]{zhou2021strongmagneticfield}%
  \BibitemOpen
  \bibfield  {author} {\bibinfo {author} {\bibfnamefont {H.}~\bibnamefont
  {Zhou}}, \bibinfo {author} {\bibfnamefont {C.}~\bibnamefont {Huang}},
  \bibinfo {author} {\bibfnamefont {N.}~\bibnamefont {Wei}}, \bibinfo {author}
  {\bibfnamefont {T.}~\bibnamefont {Taniguchi}}, \bibinfo {author}
  {\bibfnamefont {K.}~\bibnamefont {Watanabe}}, \bibinfo {author}
  {\bibfnamefont {M.~P.}\ \bibnamefont {Zaletel}}, \bibinfo {author}
  {\bibfnamefont {Z.}~\bibnamefont {Papić}}, \bibinfo {author} {\bibfnamefont
  {A.~H.}\ \bibnamefont {MacDonald}}, \ and\ \bibinfo {author} {\bibfnamefont
  {A.~F.}\ \bibnamefont {Young}},\ }\href@noop {} {\enquote {\bibinfo {title}
  {Strong-magnetic-field magnon transport in monolayer graphene},}\ } (\bibinfo
  {year} {2021}),\ \Eprint {http://arxiv.org/abs/2102.01061} {arXiv:2102.01061
  [cond-mat.mes-hall]} \BibitemShut {NoStop}%
\bibitem [{\citenamefont {Wei}\ \emph {et~al.}(2017)\citenamefont {Wei},
  \citenamefont {van~der Sar}, \citenamefont {Sanchez-Yamagishi}, \citenamefont
  {Watanabe}, \citenamefont {Taniguchi}, \citenamefont {Jarillo-Herrero},
  \citenamefont {Halperin},\ and\ \citenamefont {Yacoby}}]{wei2017mach}%
  \BibitemOpen
  \bibfield  {author} {\bibinfo {author} {\bibfnamefont {D.~S.}\ \bibnamefont
  {Wei}}, \bibinfo {author} {\bibfnamefont {T.}~\bibnamefont {van~der Sar}},
  \bibinfo {author} {\bibfnamefont {J.~D.}\ \bibnamefont {Sanchez-Yamagishi}},
  \bibinfo {author} {\bibfnamefont {K.}~\bibnamefont {Watanabe}}, \bibinfo
  {author} {\bibfnamefont {T.}~\bibnamefont {Taniguchi}}, \bibinfo {author}
  {\bibfnamefont {P.}~\bibnamefont {Jarillo-Herrero}}, \bibinfo {author}
  {\bibfnamefont {B.~I.}\ \bibnamefont {Halperin}}, \ and\ \bibinfo {author}
  {\bibfnamefont {A.}~\bibnamefont {Yacoby}},\ }\href {\doibase
  10.1126/sciadv.1700600} {\bibfield  {journal} {\bibinfo  {journal} {Science
  Advances}\ }\textbf {\bibinfo {volume} {3}},\ \bibinfo {pages} {e1700600}
  (\bibinfo {year} {2017})},\ \Eprint
  {http://arxiv.org/abs/https://www.science.org/doi/pdf/10.1126/sciadv.1700600}
  {https://www.science.org/doi/pdf/10.1126/sciadv.1700600} \BibitemShut
  {NoStop}%
\bibitem [{\citenamefont {Kim}\ \emph {et~al.}(2021)\citenamefont {Kim},
  \citenamefont {Schwenk}, \citenamefont {Walkup}, \citenamefont {Zeng},
  \citenamefont {Ghahari}, \citenamefont {Le}, \citenamefont {Slot},
  \citenamefont {Berwanger}, \citenamefont {Blankenship}, \citenamefont
  {Watanabe}, \citenamefont {Taniguchi}, \citenamefont {Giessibl},
  \citenamefont {Zhitenev}, \citenamefont {Dean},\ and\ \citenamefont
  {Stroscio}}]{kim2021edge}%
  \BibitemOpen
  \bibfield  {author} {\bibinfo {author} {\bibfnamefont {S.}~\bibnamefont
  {Kim}}, \bibinfo {author} {\bibfnamefont {J.}~\bibnamefont {Schwenk}},
  \bibinfo {author} {\bibfnamefont {D.}~\bibnamefont {Walkup}}, \bibinfo
  {author} {\bibfnamefont {Y.}~\bibnamefont {Zeng}}, \bibinfo {author}
  {\bibfnamefont {F.}~\bibnamefont {Ghahari}}, \bibinfo {author} {\bibfnamefont
  {S.~T.}\ \bibnamefont {Le}}, \bibinfo {author} {\bibfnamefont {M.~R.}\
  \bibnamefont {Slot}}, \bibinfo {author} {\bibfnamefont {J.}~\bibnamefont
  {Berwanger}}, \bibinfo {author} {\bibfnamefont {S.~R.}\ \bibnamefont
  {Blankenship}}, \bibinfo {author} {\bibfnamefont {K.}~\bibnamefont
  {Watanabe}}, \bibinfo {author} {\bibfnamefont {T.}~\bibnamefont {Taniguchi}},
  \bibinfo {author} {\bibfnamefont {F.~J.}\ \bibnamefont {Giessibl}}, \bibinfo
  {author} {\bibfnamefont {N.~B.}\ \bibnamefont {Zhitenev}}, \bibinfo {author}
  {\bibfnamefont {C.~R.}\ \bibnamefont {Dean}}, \ and\ \bibinfo {author}
  {\bibfnamefont {J.~A.}\ \bibnamefont {Stroscio}},\ }\href {\doibase
  10.1038/s41467-021-22886-7} {\bibfield  {journal} {\bibinfo  {journal}
  {Nature Communications}\ }\textbf {\bibinfo {volume} {12}},\ \bibinfo {pages}
  {2852} (\bibinfo {year} {2021})}\BibitemShut {NoStop}%
\bibitem [{\citenamefont {Paul}\ \emph {et~al.}(2022)\citenamefont {Paul},
  \citenamefont {Sahu}, \citenamefont {Watanabe}, \citenamefont {Taniguchi},
  \citenamefont {Jain}, \citenamefont {Murthy},\ and\ \citenamefont
  {Das}}]{paul2022electrically}%
  \BibitemOpen
  \bibfield  {author} {\bibinfo {author} {\bibfnamefont {A.~K.}\ \bibnamefont
  {Paul}}, \bibinfo {author} {\bibfnamefont {M.~R.}\ \bibnamefont {Sahu}},
  \bibinfo {author} {\bibfnamefont {K.}~\bibnamefont {Watanabe}}, \bibinfo
  {author} {\bibfnamefont {T.}~\bibnamefont {Taniguchi}}, \bibinfo {author}
  {\bibfnamefont {J.}~\bibnamefont {Jain}}, \bibinfo {author} {\bibfnamefont
  {G.}~\bibnamefont {Murthy}}, \ and\ \bibinfo {author} {\bibfnamefont
  {A.}~\bibnamefont {Das}},\ }\href@noop {} {\bibfield  {journal} {\bibinfo
  {journal} {arXiv preprint arXiv:2205.00710}\ } (\bibinfo {year}
  {2022})}\BibitemShut {NoStop}%
\bibitem [{\citenamefont {Zhang}\ \emph {et~al.}(2006)\citenamefont {Zhang},
  \citenamefont {Jiang}, \citenamefont {Small}, \citenamefont {Purewal},
  \citenamefont {Tan}, \citenamefont {Fazlollahi}, \citenamefont {Chudow},
  \citenamefont {Jaszczak}, \citenamefont {Stormer},\ and\ \citenamefont
  {Kim}}]{zhang2006landau}%
  \BibitemOpen
  \bibfield  {author} {\bibinfo {author} {\bibfnamefont {Y.}~\bibnamefont
  {Zhang}}, \bibinfo {author} {\bibfnamefont {Z.}~\bibnamefont {Jiang}},
  \bibinfo {author} {\bibfnamefont {J.~P.}\ \bibnamefont {Small}}, \bibinfo
  {author} {\bibfnamefont {M.~S.}\ \bibnamefont {Purewal}}, \bibinfo {author}
  {\bibfnamefont {Y.-W.}\ \bibnamefont {Tan}}, \bibinfo {author} {\bibfnamefont
  {M.}~\bibnamefont {Fazlollahi}}, \bibinfo {author} {\bibfnamefont {J.~D.}\
  \bibnamefont {Chudow}}, \bibinfo {author} {\bibfnamefont {J.~A.}\
  \bibnamefont {Jaszczak}}, \bibinfo {author} {\bibfnamefont {H.~L.}\
  \bibnamefont {Stormer}}, \ and\ \bibinfo {author} {\bibfnamefont
  {P.}~\bibnamefont {Kim}},\ }\href {\doibase 10.1103/PhysRevLett.96.136806}
  {\bibfield  {journal} {\bibinfo  {journal} {Phys. Rev. Lett.}\ }\textbf
  {\bibinfo {volume} {96}},\ \bibinfo {pages} {136806} (\bibinfo {year}
  {2006})}\BibitemShut {NoStop}%
\bibitem [{\citenamefont {Jiang}\ \emph {et~al.}(2007)\citenamefont {Jiang},
  \citenamefont {Zhang}, \citenamefont {Stormer},\ and\ \citenamefont
  {Kim}}]{jiang2007quantum}%
  \BibitemOpen
  \bibfield  {author} {\bibinfo {author} {\bibfnamefont {Z.}~\bibnamefont
  {Jiang}}, \bibinfo {author} {\bibfnamefont {Y.}~\bibnamefont {Zhang}},
  \bibinfo {author} {\bibfnamefont {H.~L.}\ \bibnamefont {Stormer}}, \ and\
  \bibinfo {author} {\bibfnamefont {P.}~\bibnamefont {Kim}},\ }\href {\doibase
  10.1103/PhysRevLett.99.106802} {\bibfield  {journal} {\bibinfo  {journal}
  {Phys. Rev. Lett.}\ }\textbf {\bibinfo {volume} {99}},\ \bibinfo {pages}
  {106802} (\bibinfo {year} {2007})}\BibitemShut {NoStop}%
\bibitem [{\citenamefont {Young}\ \emph {et~al.}(2012)\citenamefont {Young},
  \citenamefont {Dean}, \citenamefont {Wang}, \citenamefont {Ren},
  \citenamefont {Cadden-Zimansky}, \citenamefont {Watanabe}, \citenamefont
  {Taniguchi}, \citenamefont {Hone}, \citenamefont {Shepard},\ and\
  \citenamefont {Kim}}]{Young_2012_NP}%
  \BibitemOpen
  \bibfield  {author} {\bibinfo {author} {\bibfnamefont {A.~F.}\ \bibnamefont
  {Young}}, \bibinfo {author} {\bibfnamefont {C.~R.}\ \bibnamefont {Dean}},
  \bibinfo {author} {\bibfnamefont {L.}~\bibnamefont {Wang}}, \bibinfo {author}
  {\bibfnamefont {H.}~\bibnamefont {Ren}}, \bibinfo {author} {\bibfnamefont
  {P.}~\bibnamefont {Cadden-Zimansky}}, \bibinfo {author} {\bibfnamefont
  {K.}~\bibnamefont {Watanabe}}, \bibinfo {author} {\bibfnamefont
  {T.}~\bibnamefont {Taniguchi}}, \bibinfo {author} {\bibfnamefont
  {J.}~\bibnamefont {Hone}}, \bibinfo {author} {\bibfnamefont {K.~L.}\
  \bibnamefont {Shepard}}, \ and\ \bibinfo {author} {\bibfnamefont
  {P.}~\bibnamefont {Kim}},\ }\href {\doibase 10.1038/nphys2307} {\bibfield
  {journal} {\bibinfo  {journal} {Nature Physics}\ }\textbf {\bibinfo {volume}
  {8}},\ \bibinfo {pages} {550} (\bibinfo {year} {2012})}\BibitemShut {NoStop}%
\bibitem [{\citenamefont {Maher}\ \emph {et~al.}(2013)\citenamefont {Maher},
  \citenamefont {Dean}, \citenamefont {Young}, \citenamefont {Taniguchi},
  \citenamefont {Watanabe}, \citenamefont {Shepard}, \citenamefont {Hone},\
  and\ \citenamefont {Kim}}]{Maher_2013_NP}%
  \BibitemOpen
  \bibfield  {author} {\bibinfo {author} {\bibfnamefont {P.}~\bibnamefont
  {Maher}}, \bibinfo {author} {\bibfnamefont {C.~R.}\ \bibnamefont {Dean}},
  \bibinfo {author} {\bibfnamefont {A.~F.}\ \bibnamefont {Young}}, \bibinfo
  {author} {\bibfnamefont {T.}~\bibnamefont {Taniguchi}}, \bibinfo {author}
  {\bibfnamefont {K.}~\bibnamefont {Watanabe}}, \bibinfo {author}
  {\bibfnamefont {K.~L.}\ \bibnamefont {Shepard}}, \bibinfo {author}
  {\bibfnamefont {J.}~\bibnamefont {Hone}}, \ and\ \bibinfo {author}
  {\bibfnamefont {P.}~\bibnamefont {Kim}},\ }\href {\doibase 10.1038/nphys2528}
  {\bibfield  {journal} {\bibinfo  {journal} {Nature Physics}\ }\textbf
  {\bibinfo {volume} {9}},\ \bibinfo {pages} {154} (\bibinfo {year}
  {2013})}\BibitemShut {NoStop}%
\bibitem [{\citenamefont {Young}\ \emph
  {et~al.}(2014{\natexlab{a}})\citenamefont {Young}, \citenamefont
  {Sanchez-Yamagishi}, \citenamefont {Hunt}, \citenamefont {Choi},
  \citenamefont {Watanabe}, \citenamefont {Taniguchi}, \citenamefont
  {Ashoori},\ and\ \citenamefont {Jarillo-Herrero}}]{Young_2014_Nature}%
  \BibitemOpen
  \bibfield  {author} {\bibinfo {author} {\bibfnamefont {A.~F.}\ \bibnamefont
  {Young}}, \bibinfo {author} {\bibfnamefont {J.~D.}\ \bibnamefont
  {Sanchez-Yamagishi}}, \bibinfo {author} {\bibfnamefont {B.}~\bibnamefont
  {Hunt}}, \bibinfo {author} {\bibfnamefont {S.~H.}\ \bibnamefont {Choi}},
  \bibinfo {author} {\bibfnamefont {K.}~\bibnamefont {Watanabe}}, \bibinfo
  {author} {\bibfnamefont {T.}~\bibnamefont {Taniguchi}}, \bibinfo {author}
  {\bibfnamefont {R.~C.}\ \bibnamefont {Ashoori}}, \ and\ \bibinfo {author}
  {\bibfnamefont {P.}~\bibnamefont {Jarillo-Herrero}},\ }\href {\doibase
  10.1038/nature12800} {\bibfield  {journal} {\bibinfo  {journal} {Nature}\
  }\textbf {\bibinfo {volume} {505}},\ \bibinfo {pages} {528} (\bibinfo {year}
  {2014}{\natexlab{a}})}\BibitemShut {NoStop}%
\bibitem [{\citenamefont {Veyrat}\ \emph {et~al.}(2020)\citenamefont {Veyrat},
  \citenamefont {Déprez}, \citenamefont {Coissard}, \citenamefont {Li},
  \citenamefont {Gay}, \citenamefont {Watanabe}, \citenamefont {Taniguchi},
  \citenamefont {Han}, \citenamefont {Piot}, \citenamefont {Sellier},\ and\
  \citenamefont {Sacépé}}]{Veyrat_2020_Science}%
  \BibitemOpen
  \bibfield  {author} {\bibinfo {author} {\bibfnamefont {L.}~\bibnamefont
  {Veyrat}}, \bibinfo {author} {\bibfnamefont {C.}~\bibnamefont {Déprez}},
  \bibinfo {author} {\bibfnamefont {A.}~\bibnamefont {Coissard}}, \bibinfo
  {author} {\bibfnamefont {X.}~\bibnamefont {Li}}, \bibinfo {author}
  {\bibfnamefont {F.}~\bibnamefont {Gay}}, \bibinfo {author} {\bibfnamefont
  {K.}~\bibnamefont {Watanabe}}, \bibinfo {author} {\bibfnamefont
  {T.}~\bibnamefont {Taniguchi}}, \bibinfo {author} {\bibfnamefont
  {Z.}~\bibnamefont {Han}}, \bibinfo {author} {\bibfnamefont {B.~A.}\
  \bibnamefont {Piot}}, \bibinfo {author} {\bibfnamefont {H.}~\bibnamefont
  {Sellier}}, \ and\ \bibinfo {author} {\bibfnamefont {B.}~\bibnamefont
  {Sacépé}},\ }\href {\doibase 10.1126/science.aax8201} {\bibfield  {journal}
  {\bibinfo  {journal} {Science}\ }\textbf {\bibinfo {volume} {367}},\ \bibinfo
  {pages} {781} (\bibinfo {year} {2020})},\ \Eprint
  {http://arxiv.org/abs/https://www.science.org/doi/pdf/10.1126/science.aax8201}
  {https://www.science.org/doi/pdf/10.1126/science.aax8201} \BibitemShut
  {NoStop}%
\bibitem [{\citenamefont {Sodemann}\ and\ \citenamefont
  {MacDonald}(2014)}]{MacDonald_2014_FQH}%
  \BibitemOpen
  \bibfield  {author} {\bibinfo {author} {\bibfnamefont {I.}~\bibnamefont
  {Sodemann}}\ and\ \bibinfo {author} {\bibfnamefont {A.~H.}\ \bibnamefont
  {MacDonald}},\ }\href {\doibase 10.1103/PhysRevLett.112.126804} {\bibfield
  {journal} {\bibinfo  {journal} {Phys. Rev. Lett.}\ }\textbf {\bibinfo
  {volume} {112}},\ \bibinfo {pages} {126804} (\bibinfo {year}
  {2014})}\BibitemShut {NoStop}%
\bibitem [{\citenamefont {Abanin}\ \emph {et~al.}(2013)\citenamefont {Abanin},
  \citenamefont {Feldman}, \citenamefont {Yacoby},\ and\ \citenamefont
  {Halperin}}]{Halperin_2013_PRB}%
  \BibitemOpen
  \bibfield  {author} {\bibinfo {author} {\bibfnamefont {D.~A.}\ \bibnamefont
  {Abanin}}, \bibinfo {author} {\bibfnamefont {B.~E.}\ \bibnamefont {Feldman}},
  \bibinfo {author} {\bibfnamefont {A.}~\bibnamefont {Yacoby}}, \ and\ \bibinfo
  {author} {\bibfnamefont {B.~I.}\ \bibnamefont {Halperin}},\ }\href {\doibase
  10.1103/PhysRevB.88.115407} {\bibfield  {journal} {\bibinfo  {journal} {Phys.
  Rev. B}\ }\textbf {\bibinfo {volume} {88}},\ \bibinfo {pages} {115407}
  (\bibinfo {year} {2013})}\BibitemShut {NoStop}%
\bibitem [{\citenamefont {Li}\ \emph {et~al.}(2019)\citenamefont {Li},
  \citenamefont {Zhang}, \citenamefont {Yin},\ and\ \citenamefont
  {He}}]{Li_2019_PRB}%
  \BibitemOpen
  \bibfield  {author} {\bibinfo {author} {\bibfnamefont {S.-Y.}\ \bibnamefont
  {Li}}, \bibinfo {author} {\bibfnamefont {Y.}~\bibnamefont {Zhang}}, \bibinfo
  {author} {\bibfnamefont {L.-J.}\ \bibnamefont {Yin}}, \ and\ \bibinfo
  {author} {\bibfnamefont {L.}~\bibnamefont {He}},\ }\href {\doibase
  10.1103/PhysRevB.100.085437} {\bibfield  {journal} {\bibinfo  {journal}
  {Phys. Rev. B}\ }\textbf {\bibinfo {volume} {100}},\ \bibinfo {pages}
  {085437} (\bibinfo {year} {2019})}\BibitemShut {NoStop}%
\bibitem [{\citenamefont {Liu}\ \emph {et~al.}(2022)\citenamefont {Liu},
  \citenamefont {Farahi}, \citenamefont {Chiu}, \citenamefont {Papic},
  \citenamefont {Watanabe}, \citenamefont {Taniguchi}, \citenamefont
  {Zaletel},\ and\ \citenamefont {Yazdani}}]{Xiaomeng_Liu_2022}%
  \BibitemOpen
  \bibfield  {author} {\bibinfo {author} {\bibfnamefont {X.}~\bibnamefont
  {Liu}}, \bibinfo {author} {\bibfnamefont {G.}~\bibnamefont {Farahi}},
  \bibinfo {author} {\bibfnamefont {C.-L.}\ \bibnamefont {Chiu}}, \bibinfo
  {author} {\bibfnamefont {Z.}~\bibnamefont {Papic}}, \bibinfo {author}
  {\bibfnamefont {K.}~\bibnamefont {Watanabe}}, \bibinfo {author}
  {\bibfnamefont {T.}~\bibnamefont {Taniguchi}}, \bibinfo {author}
  {\bibfnamefont {M.~P.}\ \bibnamefont {Zaletel}}, \ and\ \bibinfo {author}
  {\bibfnamefont {A.}~\bibnamefont {Yazdani}},\ }\href {\doibase
  10.1126/science.abm3770} {\bibfield  {journal} {\bibinfo  {journal}
  {Science}\ }\textbf {\bibinfo {volume} {375}},\ \bibinfo {pages} {321}
  (\bibinfo {year} {2022})},\ \Eprint
  {http://arxiv.org/abs/https://www.science.org/doi/pdf/10.1126/science.abm3770}
  {https://www.science.org/doi/pdf/10.1126/science.abm3770} \BibitemShut
  {NoStop}%
\bibitem [{\citenamefont {Coissard}\ \emph {et~al.}(2021)\citenamefont
  {Coissard}, \citenamefont {Wander}, \citenamefont {Vignaud}, \citenamefont
  {Grushin}, \citenamefont {Repellin}, \citenamefont {Watanabe}, \citenamefont
  {Taniguchi}, \citenamefont {Gay}, \citenamefont {Winkelmann}, \citenamefont
  {Courtois}, \citenamefont {Sellier},\ and\ \citenamefont
  {Sacépé}}]{Coissard_2021}%
  \BibitemOpen
  \bibfield  {author} {\bibinfo {author} {\bibfnamefont {A.}~\bibnamefont
  {Coissard}}, \bibinfo {author} {\bibfnamefont {D.}~\bibnamefont {Wander}},
  \bibinfo {author} {\bibfnamefont {H.}~\bibnamefont {Vignaud}}, \bibinfo
  {author} {\bibfnamefont {A.~G.}\ \bibnamefont {Grushin}}, \bibinfo {author}
  {\bibfnamefont {C.}~\bibnamefont {Repellin}}, \bibinfo {author}
  {\bibfnamefont {K.}~\bibnamefont {Watanabe}}, \bibinfo {author}
  {\bibfnamefont {T.}~\bibnamefont {Taniguchi}}, \bibinfo {author}
  {\bibfnamefont {F.}~\bibnamefont {Gay}}, \bibinfo {author} {\bibfnamefont
  {C.}~\bibnamefont {Winkelmann}}, \bibinfo {author} {\bibfnamefont
  {H.}~\bibnamefont {Courtois}}, \bibinfo {author} {\bibfnamefont
  {H.}~\bibnamefont {Sellier}}, \ and\ \bibinfo {author} {\bibfnamefont
  {B.}~\bibnamefont {Sacépé}},\ }\href {\doibase 10.48550/ARXIV.2110.02811}
  {\enquote {\bibinfo {title} {Imaging tunable quantum hall broken-symmetry
  orders in charge-neutral graphene},}\ } (\bibinfo {year} {2021})\BibitemShut
  {NoStop}%
\bibitem [{\citenamefont {Abanin}\ \emph {et~al.}(2006)\citenamefont {Abanin},
  \citenamefont {Lee},\ and\ \citenamefont {Levitov}}]{Lee_06_Edge}%
  \BibitemOpen
  \bibfield  {author} {\bibinfo {author} {\bibfnamefont {D.~A.}\ \bibnamefont
  {Abanin}}, \bibinfo {author} {\bibfnamefont {P.~A.}\ \bibnamefont {Lee}}, \
  and\ \bibinfo {author} {\bibfnamefont {L.~S.}\ \bibnamefont {Levitov}},\
  }\href {\doibase 10.1103/PhysRevLett.96.176803} {\bibfield  {journal}
  {\bibinfo  {journal} {Phys. Rev. Lett.}\ }\textbf {\bibinfo {volume} {96}},\
  \bibinfo {pages} {176803} (\bibinfo {year} {2006})}\BibitemShut {NoStop}%
\bibitem [{\citenamefont {Shimshoni}\ \emph {et~al.}(2009)\citenamefont
  {Shimshoni}, \citenamefont {Fertig},\ and\ \citenamefont
  {Pai}}]{PRL.102.206408}%
  \BibitemOpen
  \bibfield  {author} {\bibinfo {author} {\bibfnamefont {E.}~\bibnamefont
  {Shimshoni}}, \bibinfo {author} {\bibfnamefont {H.~A.}\ \bibnamefont
  {Fertig}}, \ and\ \bibinfo {author} {\bibfnamefont {G.~V.}\ \bibnamefont
  {Pai}},\ }\href {\doibase 10.1103/PhysRevLett.102.206408} {\bibfield
  {journal} {\bibinfo  {journal} {Phys. Rev. Lett.}\ }\textbf {\bibinfo
  {volume} {102}},\ \bibinfo {pages} {206408} (\bibinfo {year}
  {2009})}\BibitemShut {NoStop}%
\bibitem [{\citenamefont
  {Kharitonov}(2012{\natexlab{a}})}]{Kharitonov_12_Edge}%
  \BibitemOpen
  \bibfield  {author} {\bibinfo {author} {\bibfnamefont {M.}~\bibnamefont
  {Kharitonov}},\ }\href {\doibase 10.1103/PhysRevB.86.075450} {\bibfield
  {journal} {\bibinfo  {journal} {Phys. Rev. B}\ }\textbf {\bibinfo {volume}
  {86}},\ \bibinfo {pages} {075450} (\bibinfo {year}
  {2012}{\natexlab{a}})}\BibitemShut {NoStop}%
\bibitem [{\citenamefont {Murthy}\ \emph {et~al.}(2014)\citenamefont {Murthy},
  \citenamefont {Shimshoni},\ and\ \citenamefont {Fertig}}]{PRB.90.241410}%
  \BibitemOpen
  \bibfield  {author} {\bibinfo {author} {\bibfnamefont {G.}~\bibnamefont
  {Murthy}}, \bibinfo {author} {\bibfnamefont {E.}~\bibnamefont {Shimshoni}}, \
  and\ \bibinfo {author} {\bibfnamefont {H.~A.}\ \bibnamefont {Fertig}},\
  }\href {\doibase 10.1103/PhysRevB.90.241410} {\bibfield  {journal} {\bibinfo
  {journal} {Phys. Rev. B}\ }\textbf {\bibinfo {volume} {90}},\ \bibinfo
  {pages} {241410} (\bibinfo {year} {2014})}\BibitemShut {NoStop}%
\bibitem [{\citenamefont {Knothe}\ and\ \citenamefont
  {Jolicoeur}(2015)}]{PRB.92.165110}%
  \BibitemOpen
  \bibfield  {author} {\bibinfo {author} {\bibfnamefont {A.}~\bibnamefont
  {Knothe}}\ and\ \bibinfo {author} {\bibfnamefont {T.}~\bibnamefont
  {Jolicoeur}},\ }\href {\doibase 10.1103/PhysRevB.92.165110} {\bibfield
  {journal} {\bibinfo  {journal} {Phys. Rev. B}\ }\textbf {\bibinfo {volume}
  {92}},\ \bibinfo {pages} {165110} (\bibinfo {year} {2015})}\BibitemShut
  {NoStop}%
\bibitem [{\citenamefont {Murthy}\ \emph {et~al.}(2016)\citenamefont {Murthy},
  \citenamefont {Shimshoni},\ and\ \citenamefont {Fertig}}]{PRB.93.045105}%
  \BibitemOpen
  \bibfield  {author} {\bibinfo {author} {\bibfnamefont {G.}~\bibnamefont
  {Murthy}}, \bibinfo {author} {\bibfnamefont {E.}~\bibnamefont {Shimshoni}}, \
  and\ \bibinfo {author} {\bibfnamefont {H.~A.}\ \bibnamefont {Fertig}},\
  }\href {\doibase 10.1103/PhysRevB.93.045105} {\bibfield  {journal} {\bibinfo
  {journal} {Phys. Rev. B}\ }\textbf {\bibinfo {volume} {93}},\ \bibinfo
  {pages} {045105} (\bibinfo {year} {2016})}\BibitemShut {NoStop}%
\bibitem [{\citenamefont {Fertig}\ and\ \citenamefont
  {Brey}(2006)}]{PRL.97.116805}%
  \BibitemOpen
  \bibfield  {author} {\bibinfo {author} {\bibfnamefont {H.~A.}\ \bibnamefont
  {Fertig}}\ and\ \bibinfo {author} {\bibfnamefont {L.}~\bibnamefont {Brey}},\
  }\href {\doibase 10.1103/PhysRevLett.97.116805} {\bibfield  {journal}
  {\bibinfo  {journal} {Phys. Rev. Lett.}\ }\textbf {\bibinfo {volume} {97}},\
  \bibinfo {pages} {116805} (\bibinfo {year} {2006})}\BibitemShut {NoStop}%
\bibitem [{\citenamefont {Kharitonov}(2012{\natexlab{b}})}]{Kharitonov_PRB_12}%
  \BibitemOpen
  \bibfield  {author} {\bibinfo {author} {\bibfnamefont {M.}~\bibnamefont
  {Kharitonov}},\ }\href {\doibase 10.1103/PhysRevB.85.155439} {\bibfield
  {journal} {\bibinfo  {journal} {Phys. Rev. B}\ }\textbf {\bibinfo {volume}
  {85}},\ \bibinfo {pages} {155439} (\bibinfo {year}
  {2012}{\natexlab{b}})}\BibitemShut {NoStop}%
\bibitem [{Note1()}]{Note1}%
  \BibitemOpen
  \bibinfo {note} {See however Ref.~\cite {PRL.128.106803} for an interesting
  recent study beyond this model}\BibitemShut {NoStop}%
\bibitem [{\citenamefont {Sheng}\ \emph {et~al.}(2006)\citenamefont {Sheng},
  \citenamefont {Weng}, \citenamefont {Sheng},\ and\ \citenamefont
  {Haldane}}]{Haldane_06}%
  \BibitemOpen
  \bibfield  {author} {\bibinfo {author} {\bibfnamefont {D.~N.}\ \bibnamefont
  {Sheng}}, \bibinfo {author} {\bibfnamefont {Z.~Y.}\ \bibnamefont {Weng}},
  \bibinfo {author} {\bibfnamefont {L.}~\bibnamefont {Sheng}}, \ and\ \bibinfo
  {author} {\bibfnamefont {F.~D.~M.}\ \bibnamefont {Haldane}},\ }\href
  {\doibase 10.1103/PhysRevLett.97.036808} {\bibfield  {journal} {\bibinfo
  {journal} {Phys. Rev. Lett.}\ }\textbf {\bibinfo {volume} {97}},\ \bibinfo
  {pages} {036808} (\bibinfo {year} {2006})}\BibitemShut {NoStop}%
\bibitem [{\citenamefont {Kane}\ and\ \citenamefont
  {Mele}(2005{\natexlab{a}})}]{KaneMele05}%
  \BibitemOpen
  \bibfield  {author} {\bibinfo {author} {\bibfnamefont {C.~L.}\ \bibnamefont
  {Kane}}\ and\ \bibinfo {author} {\bibfnamefont {E.~J.}\ \bibnamefont
  {Mele}},\ }\href {\doibase 10.1103/PhysRevLett.95.226801} {\bibfield
  {journal} {\bibinfo  {journal} {Phys. Rev. Lett.}\ }\textbf {\bibinfo
  {volume} {95}},\ \bibinfo {pages} {226801} (\bibinfo {year}
  {2005}{\natexlab{a}})}\BibitemShut {NoStop}%
\bibitem [{\citenamefont {Kane}\ and\ \citenamefont
  {Mele}(2005{\natexlab{b}})}]{KaneMele05b}%
  \BibitemOpen
  \bibfield  {author} {\bibinfo {author} {\bibfnamefont {C.~L.}\ \bibnamefont
  {Kane}}\ and\ \bibinfo {author} {\bibfnamefont {E.~J.}\ \bibnamefont
  {Mele}},\ }\href {\doibase 10.1103/PhysRevLett.95.146802} {\bibfield
  {journal} {\bibinfo  {journal} {Phys. Rev. Lett.}\ }\textbf {\bibinfo
  {volume} {95}},\ \bibinfo {pages} {146802} (\bibinfo {year}
  {2005}{\natexlab{b}})}\BibitemShut {NoStop}%
\bibitem [{\citenamefont {Ippoliti}\ \emph {et~al.}(2018)\citenamefont
  {Ippoliti}, \citenamefont {Mong}, \citenamefont {Assaad},\ and\ \citenamefont
  {Zaletel}}]{Ippoliti18}%
  \BibitemOpen
  \bibfield  {author} {\bibinfo {author} {\bibfnamefont {M.}~\bibnamefont
  {Ippoliti}}, \bibinfo {author} {\bibfnamefont {R.~S.~K.}\ \bibnamefont
  {Mong}}, \bibinfo {author} {\bibfnamefont {F.~F.}\ \bibnamefont {Assaad}}, \
  and\ \bibinfo {author} {\bibfnamefont {M.~P.}\ \bibnamefont {Zaletel}},\
  }\href {\doibase 10.1103/PhysRevB.98.235108} {\bibfield  {journal} {\bibinfo
  {journal} {Phys. Rev. B}\ }\textbf {\bibinfo {volume} {98}},\ \bibinfo
  {pages} {235108} (\bibinfo {year} {2018})}\BibitemShut {NoStop}%
\bibitem [{\citenamefont {Wang}\ \emph {et~al.}(2020)\citenamefont {Wang},
  \citenamefont {Zaletel}, \citenamefont {Mong},\ and\ \citenamefont
  {Assaad}}]{zwang_2020}%
  \BibitemOpen
  \bibfield  {author} {\bibinfo {author} {\bibfnamefont {Z.}~\bibnamefont
  {Wang}}, \bibinfo {author} {\bibfnamefont {M.~P.}\ \bibnamefont {Zaletel}},
  \bibinfo {author} {\bibfnamefont {R.~S.~K.}\ \bibnamefont {Mong}}, \ and\
  \bibinfo {author} {\bibfnamefont {F.~F.}\ \bibnamefont {Assaad}},\
  }\href@noop {} {\enquote {\bibinfo {title} {Phases of the (2+1) dimensional
  so(5) non-linear sigma model with topological term},}\ } (\bibinfo {year}
  {2020}),\ \Eprint {http://arxiv.org/abs/2003.08368} {arXiv:2003.08368
  [cond-mat.str-el]} \BibitemShut {NoStop}%
\bibitem [{\citenamefont {Blankenbecler}\ \emph {et~al.}(1981)\citenamefont
  {Blankenbecler}, \citenamefont {Scalapino},\ and\ \citenamefont
  {Sugar}}]{Blankenbecler81}%
  \BibitemOpen
  \bibfield  {author} {\bibinfo {author} {\bibfnamefont {R.}~\bibnamefont
  {Blankenbecler}}, \bibinfo {author} {\bibfnamefont {D.~J.}\ \bibnamefont
  {Scalapino}}, \ and\ \bibinfo {author} {\bibfnamefont {R.~L.}\ \bibnamefont
  {Sugar}},\ }\href {\doibase 10.1103/PhysRevD.24.2278} {\bibfield  {journal}
  {\bibinfo  {journal} {Phys. Rev. D}\ }\textbf {\bibinfo {volume} {24}},\
  \bibinfo {pages} {2278} (\bibinfo {year} {1981})}\BibitemShut {NoStop}%
\bibitem [{\citenamefont {White}\ \emph {et~al.}(1989)\citenamefont {White},
  \citenamefont {Scalapino}, \citenamefont {Sugar}, \citenamefont {Loh},
  \citenamefont {Gubernatis},\ and\ \citenamefont {Scalettar}}]{White89}%
  \BibitemOpen
  \bibfield  {author} {\bibinfo {author} {\bibfnamefont {S.}~\bibnamefont
  {White}}, \bibinfo {author} {\bibfnamefont {D.}~\bibnamefont {Scalapino}},
  \bibinfo {author} {\bibfnamefont {R.}~\bibnamefont {Sugar}}, \bibinfo
  {author} {\bibfnamefont {E.}~\bibnamefont {Loh}}, \bibinfo {author}
  {\bibfnamefont {J.}~\bibnamefont {Gubernatis}}, \ and\ \bibinfo {author}
  {\bibfnamefont {R.}~\bibnamefont {Scalettar}},\ }\href {\doibase
  10.1103/PhysRevB.40.506} {\bibfield  {journal} {\bibinfo  {journal} {Phys.
  Rev. B}\ }\textbf {\bibinfo {volume} {40}},\ \bibinfo {pages} {506} (\bibinfo
  {year} {1989})}\BibitemShut {NoStop}%
\bibitem [{\citenamefont {Assaad}\ and\ \citenamefont
  {Evertz}(2008)}]{Assaad08_rev}%
  \BibitemOpen
  \bibfield  {author} {\bibinfo {author} {\bibfnamefont {F.}~\bibnamefont
  {Assaad}}\ and\ \bibinfo {author} {\bibfnamefont {H.}~\bibnamefont
  {Evertz}},\ }in\ \href {\doibase 10.1007/978-3-540-74686-7_10} {\emph
  {\bibinfo {booktitle} {Computational Many-Particle Physics}}},\ \bibinfo
  {series} {Lecture Notes in Physics}, Vol.\ \bibinfo {volume} {739},\ \bibinfo
  {editor} {edited by\ \bibinfo {editor} {\bibfnamefont {H.}~\bibnamefont
  {Fehske}}, \bibinfo {editor} {\bibfnamefont {R.}~\bibnamefont {Schneider}}, \
  and\ \bibinfo {editor} {\bibfnamefont {A.}~\bibnamefont {Wei{\ss}e}}}\
  (\bibinfo  {publisher} {Springer},\ \bibinfo {address} {Berlin Heidelberg},\
  \bibinfo {year} {2008})\ pp.\ \bibinfo {pages} {277--356}\BibitemShut
  {NoStop}%
\bibitem [{\citenamefont {Collaboration}\ \emph {et~al.}(2021)\citenamefont
  {Collaboration}, \citenamefont {Assaad}, \citenamefont {Bercx}, \citenamefont
  {Goth}, \citenamefont {G\"otz}, \citenamefont {Hofmann}, \citenamefont
  {Huffman}, \citenamefont {Liu}, \citenamefont {Toldin}, \citenamefont
  {Portela},\ and\ \citenamefont {Schwab}}]{alfcollaboration2021alf}%
  \BibitemOpen
  \bibfield  {author} {\bibinfo {author} {\bibfnamefont {A.}~\bibnamefont
  {Collaboration}}, \bibinfo {author} {\bibfnamefont {F.~F.}\ \bibnamefont
  {Assaad}}, \bibinfo {author} {\bibfnamefont {M.}~\bibnamefont {Bercx}},
  \bibinfo {author} {\bibfnamefont {F.}~\bibnamefont {Goth}}, \bibinfo {author}
  {\bibfnamefont {A.}~\bibnamefont {G\"otz}}, \bibinfo {author} {\bibfnamefont
  {J.~S.}\ \bibnamefont {Hofmann}}, \bibinfo {author} {\bibfnamefont
  {E.}~\bibnamefont {Huffman}}, \bibinfo {author} {\bibfnamefont
  {Z.}~\bibnamefont {Liu}}, \bibinfo {author} {\bibfnamefont {F.~P.}\
  \bibnamefont {Toldin}}, \bibinfo {author} {\bibfnamefont {J.~S.~E.}\
  \bibnamefont {Portela}}, \ and\ \bibinfo {author} {\bibfnamefont
  {J.}~\bibnamefont {Schwab}},\ }\href@noop {} {\  (\bibinfo {year} {2021})},\
  \Eprint {http://arxiv.org/abs/2012.11914} {arXiv:2012.11914
  [cond-mat.str-el]} \BibitemShut {NoStop}%
\bibitem [{\citenamefont {Young}\ \emph
  {et~al.}(2014{\natexlab{b}})\citenamefont {Young}, \citenamefont
  {Sanchez-Yamagishi}, \citenamefont {Hunt}, \citenamefont {Choi},
  \citenamefont {Watanabe}, \citenamefont {Taniguchi}, \citenamefont
  {Ashoori},\ and\ \citenamefont {Jarillo-Herrero}}]{Watanabe_Young_Nature}%
  \BibitemOpen
  \bibfield  {author} {\bibinfo {author} {\bibfnamefont {A.~F.}\ \bibnamefont
  {Young}}, \bibinfo {author} {\bibfnamefont {J.~D.}\ \bibnamefont
  {Sanchez-Yamagishi}}, \bibinfo {author} {\bibfnamefont {B.}~\bibnamefont
  {Hunt}}, \bibinfo {author} {\bibfnamefont {S.~H.}\ \bibnamefont {Choi}},
  \bibinfo {author} {\bibfnamefont {K.}~\bibnamefont {Watanabe}}, \bibinfo
  {author} {\bibfnamefont {T.}~\bibnamefont {Taniguchi}}, \bibinfo {author}
  {\bibfnamefont {R.~C.}\ \bibnamefont {Ashoori}}, \ and\ \bibinfo {author}
  {\bibfnamefont {P.}~\bibnamefont {Jarillo-Herrero}},\ }\href {\doibase
  10.1038/nature12800} {\bibfield  {journal} {\bibinfo  {journal} {Nature}\
  }\textbf {\bibinfo {volume} {505}} (\bibinfo {year} {2014}{\natexlab{b}}),\
  10.1038/nature12800}\BibitemShut {NoStop}%
\bibitem [{\citenamefont {Feldman}\ \emph
  {et~al.}(2012{\natexlab{b}})\citenamefont {Feldman}, \citenamefont {Krauss},
  \citenamefont {Smet},\ and\ \citenamefont {Yacoby}}]{Yacoby_2012_Science}%
  \BibitemOpen
  \bibfield  {author} {\bibinfo {author} {\bibfnamefont {B.~E.}\ \bibnamefont
  {Feldman}}, \bibinfo {author} {\bibfnamefont {B.}~\bibnamefont {Krauss}},
  \bibinfo {author} {\bibfnamefont {J.~H.}\ \bibnamefont {Smet}}, \ and\
  \bibinfo {author} {\bibfnamefont {A.}~\bibnamefont {Yacoby}},\ }\href
  {\doibase 10.1126/science.1224784} {\bibfield  {journal} {\bibinfo  {journal}
  {Science}\ }\textbf {\bibinfo {volume} {337}},\ \bibinfo {pages} {1196}
  (\bibinfo {year} {2012}{\natexlab{b}})},\ \Eprint
  {http://arxiv.org/abs/https://www.science.org/doi/pdf/10.1126/science.1224784}
  {https://www.science.org/doi/pdf/10.1126/science.1224784} \BibitemShut
  {NoStop}%
\bibitem [{\citenamefont {Feldman}\ \emph
  {et~al.}(2013{\natexlab{b}})\citenamefont {Feldman}, \citenamefont {Levin},
  \citenamefont {Krauss}, \citenamefont {Abanin}, \citenamefont {Halperin},
  \citenamefont {Smet},\ and\ \citenamefont {Yacoby}}]{Yacoby_2013_PRL}%
  \BibitemOpen
  \bibfield  {author} {\bibinfo {author} {\bibfnamefont {B.~E.}\ \bibnamefont
  {Feldman}}, \bibinfo {author} {\bibfnamefont {A.~J.}\ \bibnamefont {Levin}},
  \bibinfo {author} {\bibfnamefont {B.}~\bibnamefont {Krauss}}, \bibinfo
  {author} {\bibfnamefont {D.~A.}\ \bibnamefont {Abanin}}, \bibinfo {author}
  {\bibfnamefont {B.~I.}\ \bibnamefont {Halperin}}, \bibinfo {author}
  {\bibfnamefont {J.~H.}\ \bibnamefont {Smet}}, \ and\ \bibinfo {author}
  {\bibfnamefont {A.}~\bibnamefont {Yacoby}},\ }\href {\doibase
  10.1103/PhysRevLett.111.076802} {\bibfield  {journal} {\bibinfo  {journal}
  {Phys. Rev. Lett.}\ }\textbf {\bibinfo {volume} {111}},\ \bibinfo {pages}
  {076802} (\bibinfo {year} {2013}{\natexlab{b}})}\BibitemShut {NoStop}%
\bibitem [{\citenamefont {Zibrov}\ \emph
  {et~al.}(2018{\natexlab{b}})\citenamefont {Zibrov}, \citenamefont {Spanton},
  \citenamefont {Zhou}, \citenamefont {Kometter}, \citenamefont {Taniguchi},
  \citenamefont {Watanabe},\ and\ \citenamefont {Young}}]{Watanabe_Young_FQH}%
  \BibitemOpen
  \bibfield  {author} {\bibinfo {author} {\bibfnamefont {A.~A.}\ \bibnamefont
  {Zibrov}}, \bibinfo {author} {\bibfnamefont {E.~M.}\ \bibnamefont {Spanton}},
  \bibinfo {author} {\bibfnamefont {H.}~\bibnamefont {Zhou}}, \bibinfo {author}
  {\bibfnamefont {C.}~\bibnamefont {Kometter}}, \bibinfo {author}
  {\bibfnamefont {T.}~\bibnamefont {Taniguchi}}, \bibinfo {author}
  {\bibfnamefont {K.}~\bibnamefont {Watanabe}}, \ and\ \bibinfo {author}
  {\bibfnamefont {A.~F.}\ \bibnamefont {Young}},\ }\href {\doibase
  10.1038/s41567-018-0190-0} {\bibfield  {journal} {\bibinfo  {journal} {Nature
  Physics}\ }\textbf {\bibinfo {volume} {14}} (\bibinfo {year}
  {2018}{\natexlab{b}}),\ 10.1038/s41567-018-0190-0}\BibitemShut {NoStop}%
\bibitem [{\citenamefont {Hegde}\ and\ \citenamefont
  {Villadiego}(2022)}]{hegde2022theory}%
  \BibitemOpen
  \bibfield  {author} {\bibinfo {author} {\bibfnamefont {S.~S.}\ \bibnamefont
  {Hegde}}\ and\ \bibinfo {author} {\bibfnamefont {I.~S.}\ \bibnamefont
  {Villadiego}},\ }\href@noop {} {\enquote {\bibinfo {title} {Theory of
  competing charge density wave, kekule and antiferromagnetic ordered
  fractional quantum hall states in graphene aligned with boron nitride},}\ }
  (\bibinfo {year} {2022}),\ \Eprint {http://arxiv.org/abs/2202.01796}
  {arXiv:2202.01796 [cond-mat.mes-hall]} \BibitemShut {NoStop}%
\bibitem [{\citenamefont {Sodemann}\ and\ \citenamefont
  {Fogler}(2012)}]{Sodemann_2012_RPA}%
  \BibitemOpen
  \bibfield  {author} {\bibinfo {author} {\bibfnamefont {I.}~\bibnamefont
  {Sodemann}}\ and\ \bibinfo {author} {\bibfnamefont {M.~M.}\ \bibnamefont
  {Fogler}},\ }\href {\doibase 10.1103/PhysRevB.86.115408} {\bibfield
  {journal} {\bibinfo  {journal} {Phys. Rev. B}\ }\textbf {\bibinfo {volume}
  {86}},\ \bibinfo {pages} {115408} (\bibinfo {year} {2012})}\BibitemShut
  {NoStop}%
\bibitem [{\citenamefont {Beach}\ \emph {et~al.}(2004)\citenamefont {Beach},
  \citenamefont {Lee},\ and\ \citenamefont {Monthoux}}]{Beach04}%
  \BibitemOpen
  \bibfield  {author} {\bibinfo {author} {\bibfnamefont {K.~S.~D.}\
  \bibnamefont {Beach}}, \bibinfo {author} {\bibfnamefont {P.~A.}\ \bibnamefont
  {Lee}}, \ and\ \bibinfo {author} {\bibfnamefont {P.}~\bibnamefont
  {Monthoux}},\ }\href {\doibase 10.1103/PhysRevLett.92.026401} {\bibfield
  {journal} {\bibinfo  {journal} {Phys. Rev. Lett.}\ }\textbf {\bibinfo
  {volume} {92}},\ \bibinfo {pages} {026401} (\bibinfo {year}
  {2004})}\BibitemShut {NoStop}%
\bibitem [{\citenamefont {Das}\ \emph {et~al.}(2022)\citenamefont {Das},
  \citenamefont {Kaul},\ and\ \citenamefont {Murthy}}]{PRL.128.106803}%
  \BibitemOpen
  \bibfield  {author} {\bibinfo {author} {\bibfnamefont {A.}~\bibnamefont
  {Das}}, \bibinfo {author} {\bibfnamefont {R.~K.}\ \bibnamefont {Kaul}}, \
  and\ \bibinfo {author} {\bibfnamefont {G.}~\bibnamefont {Murthy}},\ }\href
  {\doibase 10.1103/PhysRevLett.128.106803} {\bibfield  {journal} {\bibinfo
  {journal} {Phys. Rev. Lett.}\ }\textbf {\bibinfo {volume} {128}},\ \bibinfo
  {pages} {106803} (\bibinfo {year} {2022})}\BibitemShut {NoStop}%
\end{thebibliography}%

\clearpage

\section{Supplemental material}\label{Sec:MF}

\subsection{A. Quantum Monte Carlo implementation and examination of bulk phase transition}

We summarize the QMC implementation in this section.   
The spacial coordinate $\bm{x}$ lives on a torus of size $L_x  \times L_y$.   
We choose $L_x = 4 L_y$ for the calculation.   
The fermion annihilation operators are projected onto the ZLL:   
$\hat{\psi}_a(\bm{x}) = \sum_{n_k=1}^{N_{\phi}}  
\phi_{n_k} (\bm{x})   \hat{c}_{a, n_k }$.  
The index $a$  runs over  $1 ... 4$ corresponding to the four 
flavors constructed via physical spin and graphene sublattice.   
The wave functions for  the  ZLL,  $\phi_{n_k} (\ve{x})$,   
are defined in the Landau gauge with    
translational invariance along $y$ direction. 
$y$ direction momentum $k_y$ is a good quantum 
number ($k_y=\frac{2\pi n_k}{L_y }$).    
To implement QMC, we rewrite the various generalized pseudo-spin  
density interactions in momentum space, as follows:  
\begin{equation}\label{Eq:density_Momentum}
\begin{aligned}
 & \hat{\ve{ \psi}}^{\dagger} (\bm{x}) O^i \hat{\ve{\psi }} (\bm{x}) = \frac{1}{V} \sum_{\ve{q}} e^{-i \ve{q} \cdot \ve{x} } \hat{n}^i(\ve{q}) \\  
& \hat{ n }^i_{ \bm q }  
   =    \sum_{ k }  
     f(\bm{q} )   
     e^{ \frac{i}{2} (2 k - q_y) l_B^2 q_x }    
     [ \hat c^{\dagger}_{ k}  { O}^i  \hat c_{ k - q_y
 -  [ \frac{ k-q_y}{ 2\pi N_\phi /L_y } ]  \frac{2\pi N_\phi}{L_y} }  \\  
& - 2\delta_{q_y,0}\delta_{i,0}  ]    
   \qquad   f(\bm q)  = e^{ - \frac{1}{4} l_B^2 |\bm{q}|^2 }   
\end{aligned}
\end{equation}
where $O^i$ denote  $4 \times 4$ matrices in valley space  
$ \tau^z, \tau^x, \tau^y $ and $ \mathbb{1} $ for $i=1,2,3,0$,  
respectively.   
A complex Hubbard-Stratonovich transformation is performed to  
decouple the interaction operators for each momentum $\bm{q}$.  
The absence of the sign problem is guaranteed by    
two 
anti-unitary particle-hole symmetries $ \tau^z \sigma^x P $ and 
$ \tau^z \sigma^y P $ 
( $ P \alpha 
\hat{c}_{k} P^{-1} \equiv \Bar{\alpha } \hat{c}^{\dagger}_k $ )    
that  anti-commute with each other. 
As shown in Ref.~\cite{zwang_2020}, difficulties of this simulation  
arise from the compactness  
of the projected density operators 
in single particle Hilbert space as well as from  
the non-local commutation relation between them.   
Hence the systematic Trotter error scales maximally as $\Delta \tau^2
N_\phi^2 $,~\cite{zwang_2020} and the CPU time scales as    
$ \beta N_\phi^5 $. 
Regularization in momentum space is based on a 
truncation from the exponential energy factor  in 
density operator such that terms with $ f(\bm{q}) < 10^{-2} $ is omitted. 
On the other hand, the simulation is less efficient than the one in Ref.~\cite{zwang_2020}   
due to the breaking of $SU(2)$ spin symmetry by a finite Zeeman coupling.

Switching on a finite in-plane magnetic field,    
we reproduce the CAF-FM bulk phase diagram as a function of 
$ h  $.    
To detect the $U(1)$ symmetry breaking of the CAF state,   
we compute the  order parameter correlation function 
\begin{equation}\label{Eq:def_CAF} 
  S(\bm{q})_{\text{CAF}}  =   \frac{1}{ N_{\phi}}   
 \langle \hat{m}_{\bm{q}}  \hat{m}_{-\bm{q}}  \rangle.   
 \qquad    
\end{equation}  
Here  
$ \hat{m} (\bm{q})$ is the operator from Eq.~\ref{Eq:density_Momentum}  
corresponding to $ \tau^z \sigma^x $ (Neel order parameter). 
For an ordering wave vector $\bm{Q} = 0$, 
the local moment (magnetization)  reads 
\begin{equation}
  m_{\text{CAF}} = \sqrt{\frac{1}{N_{\phi}} S(\bm{Q})  }    
\end{equation}  

\begin{figure} 
\centering
\includegraphics[width=0.45\textwidth]{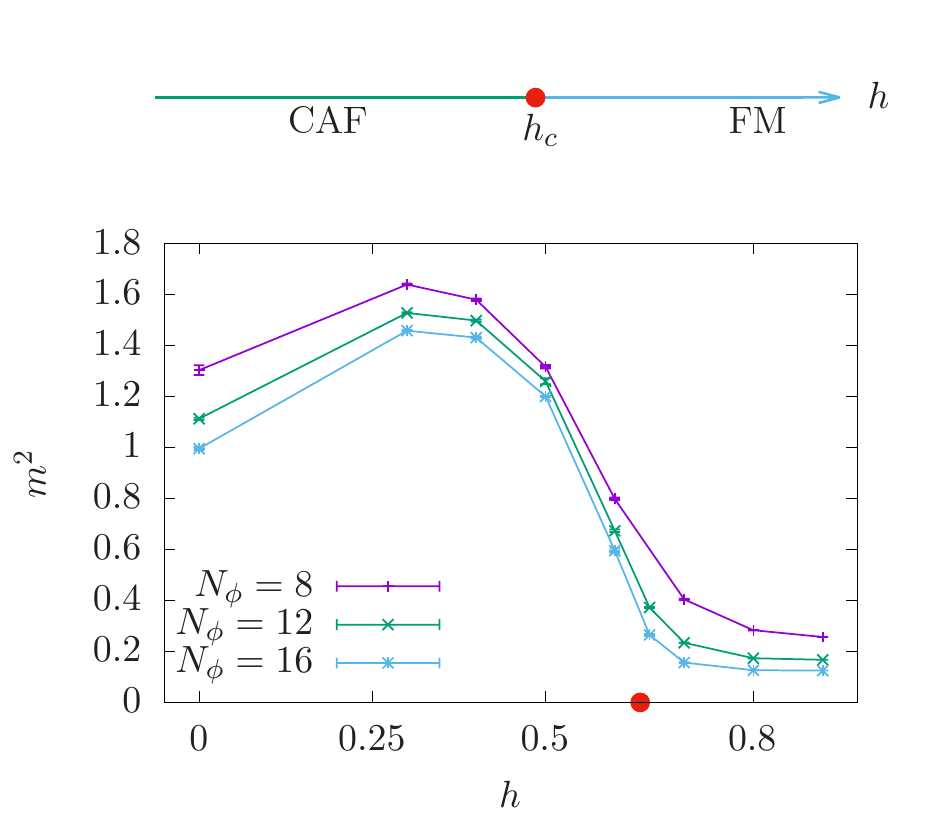} 
\caption{  
 (a) Ground state phase diagram of the bulk. 
 (b)  CAF squared magnetization $m_{CAF}$ 
 as a function of Zeeman coupling $h$.  
}\label{fig:Phase_diagram}
\end{figure}

As shown in Fig.~\ref{fig:Phase_diagram}, magnetic ordering 
develops as long as $h<h_c$, with a value of $h_c$ consistent with  
mean field theory and previous studies~\cite{Kharitonov_PRB_12}.
Here we took 
$\beta=N_\phi$ such that the simulation is converged to the ground state for each system size,  
and $ \Delta_\tau = 2 / N_\phi$.

\subsection{B. Mean field calculation}

In Hartree-Fork (HF) we use the following Slater-determinate state  
\begin{equation}
 |\Psi\rangle  \equiv \prod_{n_k=1}^{N_\phi}   
 \prod_{ a = 1 }^2  ( \sum_{b=1}^4  
  P_{a b}  \hat{c}^{\dagger}_{b, k} ) |0 \rangle   
\end{equation}
as mean field wave function.  Here the $4 \times 4$ matrix $P$ is 
defined as a projector: 
\begin{equation}
   P \equiv \frac{1}{2}  ( \mathbb{1}_4 + 
   \sin \theta \tau^z \sigma^x +  \cos \theta \sigma^z )       
\end{equation}
Hence $ \Tr P = 2 $ and $ \Psi $ is normalized. 
$\theta= \frac{\pi}{2}$ implies an AF ordering without 
canting and $\theta=0$ implies a polarized FM state.

The $g_z$ part of the interaction  
($- \frac{1}{ 2 N_{\phi}} g_z    
   \hat{n}^i_{-\bm{Q} }  \hat{n}^i_{\bm{Q} }  $) 
contributes a mean field energy of  
\begin{equation}
\begin{aligned}
  E_Z &
  = \frac{1}{ 4 \pi } g_z
  ( \Tr [ P \tau_z ]^2 - \Tr[P \tau_z P \tau_z]  ) 
  \sum_{\bm{q}} f (\bm{q})  f (-\bm{q}) \\ 
 & = - \frac{1}{4\pi} g_z (1+\cos^2 \theta+\sin^2 \theta ) N_{ \phi } \\ 
 & = - \frac{g_z}{ 2\pi }  N_{\phi} 
\end{aligned} 
\end{equation} 
where $f (\bm{q}) \equiv \exp (-\frac{1}{4} |\bm{q}|^2 l_B^2) $
is the exponentially decaying form factor of the density operator.

The $g_{\perp} $ part of the interaction contributes mean field 
energy of 
\begin{equation} 
\begin{aligned}
  E_{\perp } &
  = \frac{1}{ 4 \pi }  2 g_{\perp }  
  ( \Tr [ P \tau_x ]^2 
  - \Tr[P \tau_x  P \tau_x ]  )       
  \sum_{\bm{q}} f (\bm{q})  f (-\bm{q}) \\    
  & = - \frac{1}{ 2 \pi }  g_\perp   
   ( 1 + \cos^2 \theta - \sin^2 \theta ) N_{ \phi } \\ 
  & = -  \frac{ g_\perp}{\pi} \cos^2 \theta    N_{\phi} 
\end{aligned} 
\end{equation} 
On the other hand the $SU(4)$ invariant Coulomb interaction $g_0$ 
does not contribute to the mean field energy.

The ground state mean field energy, considering all the interactions 
as well as Zeeman coupling is then  
\begin{equation}
\begin{aligned}
   E_{MF}/N_\phi  
 & = - \frac{ g_z + 2 g_{\perp} \cos^2 \theta }{2\pi} g_z   
   - 2 h \cos \theta   \\      
 & =  \frac{- g_z + 2 | g_{\perp} | \cos^2 \theta}{ 2\pi } 
   - 2 h \cos \theta  
\end{aligned}
\end{equation}
Hence the minimal energy happens at $ \cos \theta=
\frac{ \pi h}{g_\perp}$.

And the mean field Hamiltonian reads 
\begin{equation}\label{Eq:H_MF}
  \hat{H}_{ \text{MF} } = \sum_{ n_k=1}^{N_\phi} 
  \hat{c}^{\dagger}_k h_{ \text{int} } \hat{c}_k  - 
   h \sum_{ n_k=1}^{N_\phi} 
  \hat{c}^{\dagger}_k \sigma_z \hat{c}_k
\end{equation}
where 
\begin{equation}
\begin{aligned}
   h_{ \text{int} }
  & = \frac{1}{ 2 \pi } [ \sum_{i=1}^{3} g_i 
  [ T^i \Tr (P T^i) - T^i P T^i ]  -   \widetilde{g_0}  P ]  \\  
  & = \frac{1}{4\pi} ( -  \widetilde{g_0} - g_z + 2 g_{\perp} ) 
   \sin \theta   \tau^z \sigma^x   \\ 
  & +   \frac{1}{4\pi} ( -  \widetilde{g_0} - g_z - 2 g_{\perp} )
  \cos \theta    \sigma^z  \\ 
  & = \frac{1}{4\pi} ( -  \widetilde{g_0} - g_z - 2|g_{\perp}|) 
  \sin \theta  \tau^z \sigma^x   \\ 
  & + \frac{1}{4\pi} ( -  \widetilde{g_0} - g_z + 2|g_{\perp}|) 
  \cos \theta  \sigma^z        
\end{aligned}  
\end{equation}  
where mean field contribution  from  
Coulomb interactions is  
\begin{equation}
   \widetilde{g_0} =   \frac{e^2}{ \epsilon l_B }
   \sum_{\bm{q}} 
  \frac{f(\bm{q}) f(-\bm{q})    
   V_0 (\bm{q}) }{N_\phi}  
\end{equation} 
The momentum dependent potential is the Fourier component of 
$ V(| \bm{x}-\bm{x}' |) $ in main text: 
\begin{equation}
  V_0 (\bm{q}) = \frac{1}{ |\bm{q}| l_B } ( 1-e^{ -|\bm{q}| d } )     
\end{equation}

Due to the anti-commutation relation between $\sigma^z$ and 
$ \tau^z \sigma^x $, the one particle
excitation gap of Eq.~\ref{Eq:H_MF} is 
\begin{equation}\label{Eq:Gap_HF} 
\begin{aligned}
 \Delta_{\text{sp}}  
=&  \{ [\frac{1}{4\pi}( \widetilde{g_0} + g_z - 2|g_\perp|) \cos \theta +h ]^2   \\ 
+&   [\frac{1}{4\pi}( \widetilde{g_0} + g_z + 2|g_\perp|) \sin \theta ]^2 \}^{1/2} 
\end{aligned}
\end{equation}  
where the phase angle is given by 
$ \cos \theta= {\pi h} / { g_\perp}$.

\subsection{C. Finite size dependence of single particle gap}

\begin{figure}
\centering
\includegraphics[width=0.49\textwidth]{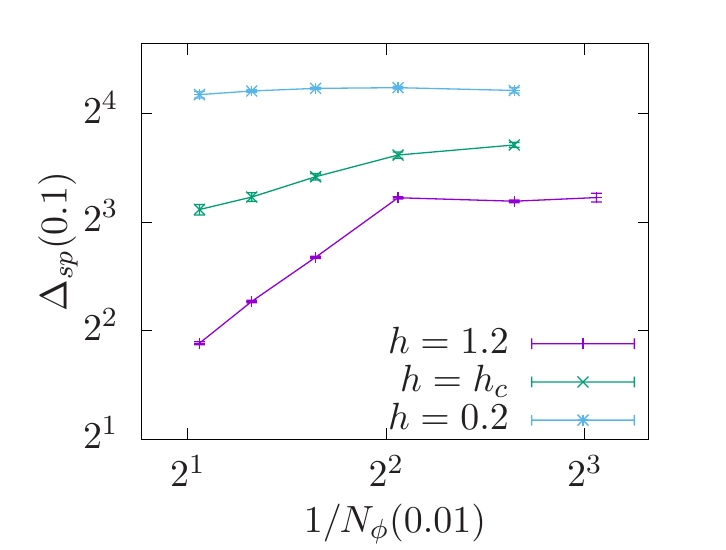} 
\caption{  
 $1/N_\phi$ dependence of $\Delta_{sp}$ in log-log scale. 
}\label{fig:Gap_N} 
\end{figure}

In this section we show the finite size dependence of single particle excitation gap 
at the edge. 
We extrapolate the gap $\Delta_{\text{sp}}$ asymptotically based on: 
\begin{equation}
    \sum_{a} \langle 
   \hat{\psi}^{\dagger}_{a } (x, y, \tau) 
   \hat{\psi}_{a } (x, y,  0 )   
   \rangle     \propto 
   e^{ - \Delta_{\text{sp}}(x) \tau }    
\end{equation}  
for large $\tau$.  We take edge gap as
$ \Delta_{\text{sp}} \equiv \min \{ \Delta_{\text{sp}}(x) \}$.   
Fig.~\ref{fig:Gap_N} displays the system size dependence of 
$\Delta_{\text{sp}}$ for three values of Zeeman coupling.  
A linear behavior on a doubly logarithmic scale indicates vanishing value of  
$\Delta_{\text{sp}}$ in the thermodynamic limit  for the case of 
$h=1.2$ and $h_c=2 / \pi $.    

\subsection{D. Vanishing long range Coulomb interaction case}

\begin{figure*}
\centering
\includegraphics[width=0.97\textwidth]{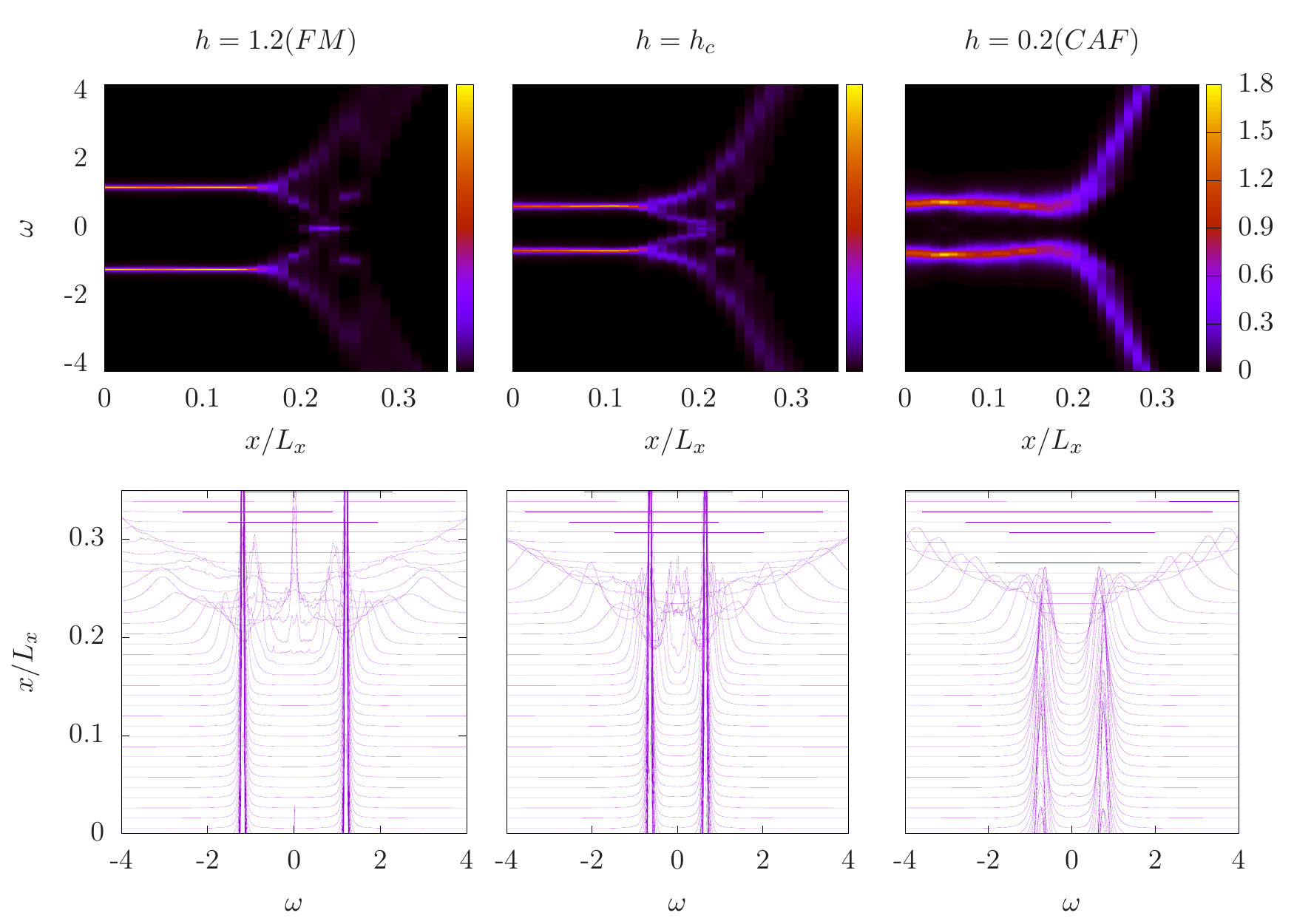} 
\caption{  
 Same as Fig.~\ref{fig:Green_edge} in main text, 
 for $ \hat{H}_{\text{Coul}} = 0 $. 
}\label{fig:Green_edge_U0} 
\end{figure*}

We claimed in the main text that the broadening of
the quasi-particle peak at $h_c$ is mainly induced by  strong long range Coulomb 
interaction that which are SU(4) invariant.  Here we show the LDOS in the absence of  
$ \hat{H}_{\text{Coul}} $.     
As shown in Fig.~\ref{fig:Green_edge_U0},  finite size gap along the edge vanishes and
the spectral weight shows well defined quasi-particle  
behavior, which is close to the one from mean field analysis~\cite{Kharitonov_12_Edge}.  

\end{document}